\documentstyle[epsfig,12pt,epsf,cite]{article}

\newcommand{\apgt} {\ {\raise-.5ex\hbox{$\buildrel>\over\sim$}}\ }
\newcommand{\aplt} {\ {\raise-.5ex\hbox{$\buildrel<\over\sim$}}\ } 
\begin{document}

\begin{center}
{\large\bf
Formation of mesic nuclei by ($\gamma$,p) reactions
}

\vspace{3mm}

{H.~Nagahiro$^a$, D.~Jido$^b$
\footnote{ Present address: Yukawa Institute for Theoretical Physics, 
Kyoto University, Kyoto 606-8502, Japan} 
and S.~Hirenzaki$^c$}
\vspace{3mm}

{\small \em
$^a$Research Center for Nuclear Physics (RCNP), Osaka University,
 Ibaraki, Osaka 567-0047, Japan\\
$^b$Physik-Department, Technische Universit\"at M\"unchen,
D-85747 Garching, Germany\\
$^c$Department of Physics, Nara Women's University, Nara 630-8506, Japan\\
}
\end{center}

\abstract{
We present a theoretical study on formation rates of $\eta$ and $\omega$ meson-nucleus systems induced by $(\gamma,p)$ reactions on nuclear targets at ideal recoilless condition. 
We find that the smaller distortion effect in the $(\gamma,p)$ reaction
enables us to investigate properties of the mesons created deeply inside
nucleus more clearly. We also consider excitation of scalar-isoscalar $(\sigma)$ mode
in nucleus in order to investigate spectral enhancement around two-pion
threshold caused by partial restoration of chiral symmetry. We conclude
that valuable information of meson-nucleus interactions can be extracted
from global structure of the missing mass spectra in the $(\gamma,p)$
reaction. 
}

\section{Introduction}
The study of the in-medium properties of hadrons is one of the most important
subjects in contemporary nuclear physics and has attracted continuous
attention. 
The detailed investigation of hadron-nucleus bound systems clarifies
quantitative information on basic hadron-nucleus interactions. This is
one of the important steps for understanding the QCD phase structure.  
So far, atomic states of pion, kaon and $\bar p$ have been
investigated comprehensively and observed experimentally~\cite{Batty97}. 
The recent interest 
reaches the extension to the systems with other heavier neutral mesons, such as 
the 
$\eta$ and $\omega$
mesons, governed purely by strong interaction in contrast to the atomic states. 
The scalar-isoscalar ($\sigma$) excited mode in nucleus is also
considered as one of the interesting systems.

%

Recent development in experimental aspect is the establishment of the
(d,$^{3}$He) spectroscopies in the formation of the deeply bound pionic
atoms~\cite{NPA530etc,PRC62,PRL88,PLB514}.
It opens new possibilities of the formations of other hadron-nucleus
bound systems ~\cite{PRC66(02)045202, PRC68(03)035205, EPJA6,
PLB443etc}. The ($\gamma$,p) reaction is another experimental 
tool for 
formation of 
the hadron-nucleus system and originally proposed in
Refs.~\cite{PLB231,JPhysG,PLB527,PLB502}.  
The ($\gamma$,p) reaction, as well as the (d,$^{3}$He) reaction, is able to
satisfy recoilless condition at the meson creation energy. The use of the
($\gamma$,p) reaction has good advantage that the distortion effects of the
projectile and the ejectile are expected to be smaller than in the (d,$^{3}$He)
reaction mainly due to  the incident photon.

It is also important mentioning that, in the theoretical point of view, the study
of the hadron in finite density gives us information on the QCD phase
structure~\cite{PR247etc}, 
such as partial restoration of chiral symmetry, in which a reduction of the
chiral condensate takes place in nuclear density.  In this line hadronic bound
systems have been investigated in various chiral
models~\cite{APPB31etc,PRC61,PRC66(02)045202,PRC68(03)035205,PLB550, 
NPA624,NPA650,NPA710}.

In this paper, 
motivated by the successful study of the (d,$^{3}$He) reaction and the advantage of the smaller distortion effect,
we discuss the ($\gamma$,p) reaction for studying the formations of the
various meson-nucleus systems ($\eta$, $\omega$ and scalar-isoscalar
($\sigma$) mode) at the recoilless kinematics. 
Structure of the mesic nucleus is studied theoretically using the
appropriate meson-nucleus optical potentials in nuclear matter.
In order to evaluate the formation rates of the mesic nuclei, we introduce a Green function constructed by the Klein-Gordon equation with the optical potential, and calculate the cross section of the ($\gamma$,p) reaction on a
nuclear target following the method by Morimatsu and Yazaki \cite{NPA435}.
In the case of the scalar-isoscalar ($\sigma$) mode, we evaluate an
effective spectral function 
of this mode
in nucleus because of both the lack of the elementary
cross section and the large imaginary part of sigma meson self-energy in
vacuum, which requires certain modifications in the Green function
method.

In the previous paper~\cite{PRC66(02)045202,PRC68(03)035205} for studying the $\eta$-nucleus system
in the chiral doublet model,  we found repulsive nature of the $\eta$
optical potential inside the nucleus which is associated with reduction
of the mass difference of $N$ and $N(1535)$ caused by partial
restoration of chiral symmetry. Consequently there exist certain
discrepancies between the spectra obtained by the chiral doublet model
and the chiral unitary model.  
The discrepancies are expected to be
distinguished by experimental data \cite{PRC68(03)035205}.
However, in the (d,$^3$He) reaction, due to the large distortion effect,
it might be difficult 
to obtain clear information about the
optical potential in the 
nuclear center.
Therefore, it is interesting to consider the ($\gamma$,p) reaction
with much smaller distortion effects to explore the behavior of the $\eta$ meson in the nucleus.

As 
another meson-nucleus systems, we 
consider the $\omega$-mesic
nuclei. 
The $\omega$ mesic nuclei were studied by
several groups and the theoretical results have been reported 
in Refs.~\cite{EPJA6,PLB443etc,PLB502,NPA624,NPA650}.
Recently, several experimental results have been also reported in
Refs.~\cite{ne0504016,ne0504010}.
Especially, in Refs.~\cite{NPA624,NPA650,PLB502},
%
one finds the $\omega$ self-energy in 
nuclear medium
based on an effective model and the calculated ($\gamma$,p) spectra for
the production of the $\omega$-mesic nuclei at the recoilless kinematics.
%
In 
the present
paper, we calculate the ($\gamma$,p) spectra for the formation
of the $\omega$-mesic nuclei, and we show the incident $\gamma$ energy
dependence of the whole spectra and the dominant subcomponents to study
the experimental feasibilities in the lower energy photon facilities.
This study is also interesting for the higher energy photon facility
like SPring-8, since the incident photon energy has a certain
distribution and we need to use the photon with various energies around
the ideal recoilless kinematics to get better
statistics~\cite{muramatsu}.
We also study the sensitivity of the expected spectra to the
$\omega$-nucleus interaction using another theoretical
prediction~\cite{Lutz:2001mi}.

We also consider the ($\gamma$,p) reactions in the scalar-isoscalar
($\sigma$) channel
inspired by Ref.~\cite{PRD57etc, PRL82, PRD63}, where
%
an enhanced spectral function
near the $2\pi$
threshold has been 
pointed out
as a characteristic signal of the partial
restoration of 
chiral symmetry. 
This is associated with a conceivable reduction of the sigma mass and width in
medium. 
Therefore, 
there might be a chance to create the $\sigma$ mesic nuclei with relatively
small widths, if deeply bound states are formed in heavy
nuclei~\cite{NPA710}. 
It has been 
also
found 
difficult, 
in the (d,t) or (d,$^3$He) reactions, 
to observe the enhancement near the $2\pi$ threshold due to
the large distortion effects~\cite{NPA710}.  
In this paper, we explore the possibility of production of the
$\sigma$-mesic nuclei by the photon-induced reactions, expecting 
that
the `transparency' of the ($\gamma$,p) reactions enables us to observe 
the enhancement of the spectral function near the $2\pi$ threshold 
caused by the partial restoration of chiral symmetry in the medium.

%
%

This paper is organized as follows.
In Sec.\ref{sec2}, we will discuss our model for the optical potential
of each meson separately, and study the structure of the mesic nuclei
showing the binding energies and the widths obtained with the optical
potential.  
In Sec.\ref{sec3}, we 
will show 
the advantages of the ($\gamma$,p) 
reaction for 
the formation of
the meson-nucleus system, and review briefly the formulation to
calculate the formation spectrum of the mesic nucleus.   
In Sec.\ref{sec4}, we 
will present
the numerical results of the missing mass spectra of
the ($\gamma$,p) reaction, 
and compare them
with those of the (d,$^{3}$He)
reaction. Finally Sec.\ref{sec5} is devoted to summary of this paper.

\section{Structure of mesic nuclei}
\label{sec2}

In this section, we describe the theoretical formulation to study the
structure of the meson-nucleus systems and show their  
numerical results,
such as the binding energies and widths,
within an optical potential approach,
in which all the meson-nucleus interactions are summarized in an optical potential of the meson in nuclear matter. 

The wave functions, the binding
energies and their widths 
for the in-medium meson (if it is bound)
are calculated by solving the Klein-Gordon
equation with the optical potential. 
%
%
%
The Klein-Gordon equation is written as
\begin{equation}
 \left[-\nabla^2+\mu^2+2\mu V(\omega,\rho(r))\right]\phi(r)=E^2\phi(r),
\label{eq:K-Geq}
\end{equation}
where $\mu$ denotes the meson-nucleus reduced mass, which
is very close to the meson mass for heavy nuclei, 
$\omega$ is the real part of the relativistic meson energy $E$, and
$V(\omega,\rho(r))$ presents the optical potential for the
meson-nucleus system.
Here we assume to neglect the momentum dependence of the optical potential,
since we consider the recoilless condition on the ($\gamma,$p) reaction for
the mesic nuclei formation, in which the meson will be created in the nucleus
nearly at rest. 
The optical potential $V$ and the eigenenergy $E$ are 
%
expressed as complex numbers
in general
because of the absorptive 
effects of the nucleus.
We need to solve
the Klein-Gordon equation (\ref{eq:K-Geq}) in a self-consistent manner for the energy
in case 
the optical potential $V$ has the energy dependence.
We follow the method of Kwon and Tabakin to solve the Klein-Gordon
equation~\cite{PRC18}, which was successfully applied to calculate pionic
atom states~\cite{PLB194}. Here, we increase the number of mesh points
in the momentum space about 10 times larger than the original work
Ref.~\cite{PRC18} as discussed in Ref.~\cite{PRC66(02)045202}.

%

In the following parts of the present section, we discuss the in-medium
optical potential $V(\omega,\rho)$ for the $\eta$, $\omega$ and $\sigma$
mesons separately.
We show the bound state spectra, if the mesons are bound,
solving the Klein-Gordon equation (\ref{eq:K-Geq}) with the optical potential.
Here we assume the local density approximation with the nuclear density
distribution $\rho(r)$ written as the empirical Woods-Saxon form:
\begin{equation}
\rho(r)=\frac{\rho_0}{1+\exp(\frac{r-R}{a})},
\label{eq:W-S}
\end{equation}
with the radius of nucleus $R=1.18A^{1/3}-0.48$ fm and the diffuseness $a=0.5$
fm for the nuclear mass number $A$. 

\subsection{$\eta$-nucleus optical potential}
\label{sec:etapot}
First of all, we discuss the $\eta$-nucleus system.
The $\eta$-mesic nuclei were studied by Haider and Liu~\cite{HaiderLiu}
and by Chiang, Oset and Liu~\cite{PRC44(91)738}. As for the formation
reaction, the attempt to find the bound states by the ($\pi^+$,p)
reaction led to a negative result~\cite{PRL60(88)2595}. Recently,
some experiments in photoproduction processes 
indicated
observations of such bound
states in $^{12}$C target~\cite{Sokol} and $^3$He target~\cite{Pfeiffer},
and another experiment is now planed to 
investigate
the $\eta$-nucleus interaction~\cite{EPJA6}.
In this study
we use the same theoretical models for $\eta$-nucleus interaction as
described in Refs.~\cite{PRC66(02)045202,PRC68(03)035205} in 
further
detail.
%
In the $\eta$-nucleon system, the $N(1535)$ resonance
$(N^{*})$ plays an important role due to the 
dominant
$\eta NN^{*}$ coupling.
Here we evaluate
the $\eta$-nucleus optical potential $V_{\eta}(\omega,\rho(r))$
in the two different models 
which are
based on distinct physical pictures of $N^{*}$. One
is the chiral doublet model.
This 
is an extension of the linear sigma model
for the nucleon and its chiral
partner~\cite{PRD39(89)2805,PTP106(01)873etc,PRD57(98)4124}.
The other is the chiral unitary model, in which
$N^{*}$ is dynamically generated in the coupled channel 
meson-baryon scattering~\cite{PLB550,NPA612}. 

In the first approach, the $N^{*}$ is introduced as a particle with a large
width and appears in an effective Lagrangian together with the nucleon field.
Assuming $N^{*}$-hole excitation induced by the $\eta$ meson in nucleus,
we obtain
the $\eta$-nucleus optical potential at finite nuclear density
as,
\begin{equation}
V_\eta(\omega,\rho(r))
= \frac{g_\eta^2}{2\mu}\frac{\rho(r)}{\omega+m^*_N(\rho)
-m^*_{N^*}(\rho)+i\Gamma_{N^*}(\omega,\rho)/2
},
\label{eq:eta-potential}
\end{equation}
in the local density approximation and the heavy baryon
limit\cite{PRC44(91)738}. Here
$\mu$ is the $\eta$-nucleus reduced mass
and
$\rho(r)$ is the density distribution of the nucleus.
%
The $\eta NN^{*}$ coupling is assumed to be $S$-wave:
\begin{equation}
 {\cal L}_{\eta NN^*}(x)=g_\eta \bar{N}(x)\eta(x)N^*(x)+{\rm H.c.},
\label{eq:Lagrangian_etaNN}
\end{equation}
and the coupling constant $g_\eta$ is determined to be $g_\eta \simeq 2.0 $
in order to
reproduce the partial width $\Gamma_{N^*\rightarrow\eta N} \simeq 75$MeV
at tree level.
The
$S$-wave nature of the $\eta NN^{*}$ vertex simplifies the particle-hole loop
integral in Eq.~(\ref{eq:eta-potential}).
$m^*_N$ and $m^*_{N^*}$ are the effective
masses of $N$ and $N^*$ in the nuclear medium, respectively.
Considering that the $N^{*}$ mass in free space lies only 50 MeV above the
threshold and that the mass difference of $N$ and $N^{*}$ might 
change
in the medium, the $\eta$-nucleus optical potential is expected to be extremely
sensitive to the in-medium properties of $N$ and $N^{*}$. For instance, if the
mass difference reduces in the nuclear medium as $m_{\eta} + m_{N}^{*} -
m_{N^{*}}^{*} > 0$, then the optical potential turns to be repulsive
\cite{PRC66(02)045202}. 

Using the free space values of the $N$ and $N^{*}$
masses in Eq.~(\ref{eq:eta-potential}), we find that the optical potential 
gives an attractive $\eta N$ scattering length, $a_{\eta}= 0.24+i0.38$~fm,
which is comparable to $a_\eta=0.20+i0.26$~fm
and $a_\eta=0.26+i0.25$~fm as 
reported in Ref.~\cite{NPA612,Inoue}, respectively.
On the other hand, the
value of the $\eta N$ scattering length $a_\eta$ of our model is
smaller than the recent analyses, in which larger values are reported
such as $a_\eta\sim 0.9 + i0.3$ fm~\cite{NPA543,PRC71}.
In the present study, 
however, our main purpose is to study the effects of the partial
restoration of chiral symmetry in medium. As we shall see below,
in the chiral doublet scenario of $N$ and $N^*$,
the
repulsive $\eta$-nucleus interaction can be realized in the nuclear
medium due to the symmetry restoration. This exotic behavior of the
$\eta$-nucleus interaction is independent of the absolute value of
$a_\eta$ in our model. Thus, we start with the optical potential
(\ref{eq:eta-potential}) and
take into
account the effects of the chiral symmetry restoration properly to $N^*$
and the $\eta$-mesic nuclei.

The chiral doublet model is used for the calculation of the in-medium masses
of $N$ and $N^{*}$ and the in-medium $N^{*}$ width in 
Eq.~(\ref{eq:eta-potential}). The Lagrangian of the chiral doublet 
model 
(the mirror model)
is given as
\cite{PRD39(89)2805,PTP106(01)873etc}, 
\begin{eqnarray}
{\cal L} &=& \sum_{i=1,2} \left[\bar N_{i} i \partial \hspace{-6pt} / N_{i} -
   g_{i} \bar N_{i} (\sigma + (-)^{i-1} i \gamma_{5} \vec\tau\cdot \vec\pi)
N_{i}\right]\nonumber\\
&& - m_{0} (\bar N_{1}\gamma_{5}N_{2}-
\bar N_{2}\gamma_{5}N_{1}) + {\cal L}_{\rm meson}
\end{eqnarray}
where the nucleon fields $N_{1}$ and $N_{2}$ have positive and negative
parities, respectively, and the physical nucleons $N$ and $N^{*}$ are
expressed as a linear combination of $N_{1}$ and $N_{2}$. The parameters
are determined so that the model reproduces the free space values of the $N$ and
$N^{*}$ masses and the partial decay width of $N^{*}$ to $\pi N$ as
$\Gamma_{N^*\rightarrow\pi N}=75$ MeV 
with $\langle \sigma \rangle_0 = 93$ MeV
\cite{PTP106(01)873etc}. 

In the chiral doublet model together with the assumption of partial
restoration of chiral symmetry, a reduction of the mass difference of $N$ and
$N^{*}$ in the medium is found to be 
in the mean field approximation as
\cite{PRD39(89)2805,PTP106(01)873etc,PLB224(89)11,NPA640(98)77},
\begin{equation}
m^*_N(\rho)-m^*_{N^*}(\rho)=\Phi(\rho)(m_N-m_{N^*}),
\label{eq:mass_dif}
\end{equation}
where $m_{N}$ and $m_{N^{*}}$ are the $N$ and $N^{*}$ masses in free space,
respectively, and
\begin{eqnarray}
\Phi(\rho)=1-C{ \rho \over \rho_0} \ . \label{eq:defPhi}
\end{eqnarray}
Here we take the linear density approximation of the in-medium modification of
the chiral condensate, and the parameter $C$ represents the strength of the
chiral restoration at the nuclear saturation density $\rho_{0}$. The empirical
value of $C$ lies from 0.1 to 0.3~\cite{PRL82}. Here we perform our
calculations with
$C=0.0$ and $0.2$ in order to investigate the effect of the partial restoration
of chiral symmetry. A self-consistent calculation within the chiral doublet model
gives moderately linear dependence of the chiral condensate to the density 
with
$C=0.22$ in Ref.~\cite{NPA640(98)77}, in which the relativistic  Hartree
approximation is used to 
calculate the chiral condensate in nuclear matter.

\begin{figure}[hbt]
\epsfxsize=12cm
\centerline{
\epsfbox{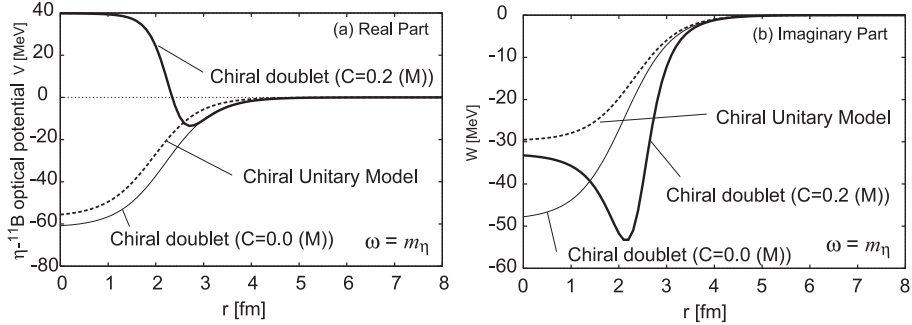}
}
\caption{
The $\eta$-nucleus optical potential for the $\eta$-$^{11}$B system as
functions of the radius coordinate $r$ reported in
 Ref.~\cite{PRC68(03)035205}. The left and right figures 
 show the real part and the imaginary part of the $\eta$-nucleus optical
 potential, respectively.
In both figures, the solid lines show the potentials of the chiral
 doublet model with $C=0.2$ (thick line) and $C=0.0$ (thin line), and 
 dashed lines show that of the chiral unitary model, which is 
picked 
 from the results shown in Ref.\cite{PLB550}
\label{fig:Vopt}
}
\end{figure}

%
%
In Fig.~\ref{fig:Vopt}, we show the $\eta$-nucleus optical potentials
obtained by the chiral doublet model with $C=0.0$ and $C=0.2$ 
%
in the case of the $\eta$-$^{11}$B system assuming the local density
approximation and the Woods-Saxon nuclear distribution defined in
Eq.~(\ref{eq:W-S}).
The potential with $C=0.0$ corresponds to the so-called $t\rho$
approximation because there are no medium modifications for $N$ and
$N^*$.
As shown in the figure, the potential with $C=0.2$ has quite different shape
from the case with $C=0.0$.
The
$\eta$-nucleus optical potential with $C=0.2$ turns to be
repulsive above a critical density $\rho_c$ at
which the sign of $\omega+m^*_N-m^*_{N^*}$ becomes positive.
Consequently, the $\eta$-nucleus optical potential has a curious shape
of
a repulsive core inside nucleus and an attractive
pocket in nuclear surface as indicated in Fig.\ref{fig:Vopt}. 
The qualitative feature of the optical potential discussed here,
such as the appearance of the repulsive core in the case of $C=0.2$,
is independent
of type of nuclei due to saturation of nuclear density. 
It is
extremely interesting to confirm the existence (or non-existence) of this
curious shaped potential experimentally.

We should mention here that there are two types of the chiral doublet
model, namely, the mirror assignment and the naive assignment,
due to two
possible type of
assignment for the axial charge to the $N$ and $N^*$
\cite{PRD39(89)2805,PTP106(01)873etc}.
The potential in Fig. \ref{fig:Vopt} is obtained by the mirror
assignment case.
In the mirror assignment, $N^*$ is regarded as the chiral partner of $N$
and forms a chiral multiplet together with $N$.
While in the naive assignment, $N$ and $N^*$ are treated as independent
baryons
in the sense of the chiral group. 
We calculate the reaction spectra using both assignments, and show the
results in next section.
In both cases, the mass difference of $N$ and $N^*$ is given
in the same form as shown in 
Eq.~(\ref{eq:mass_dif}).
Hence,
the energy and the density dependence of
the potential of the
naive assignment resemble those of the mirror assignment.
The detail discussions are given in Ref.~\cite{PRC68(03)035205}.
%

Let us move on the second approach of the chiral unitary
model~\cite{PLB550,NPA612}.
In this approach, 
the $N^{*}$ resonance is expressed as a dynamically 
generated object in the meson baryon scattering, and
one solves a coupled channel Bethe-Salpeter equation to obtain
the $\eta$-nucleon scattering amplitude.
The optical potential in medium is obtained by closing the
nucleon external lines in the $\eta N$ scattering amplitude 
and considering the in-medium effect on the
scattering amplitude, such as Pauli blocking.
Since the $N^{*}$
in the chiral unitary approach
is found to have a large component of $K\Sigma$ 
and the $\Sigma$ hyperon is free from the Pauli blocking in nuclear
medium,
very little 
mass shift of $N^{*}$ is expected in the medium
\cite{PLB362(95)23}, while the chiral doublet
model predicts the significant mass reduction 
as discussed above.
The optical potential obtained 
by
the chiral unitary approach in
Ref.~\cite{PLB550} is also shown in Fig.~\ref{fig:Vopt}, and the
potential resembles
that of the chiral doublet model with $C=0.0$
due to the small medium effect.

In Ref.\cite{PRC66(02)045202}, in order to obtain  the
eigenenergies, we have solved the Klein-Gordon equation
Eq.~(\ref{eq:K-Geq}) numerically with the $\eta$-nucleus optical
potential given in Eq.~(\ref{eq:eta-potential})
for the mirror assignment. We show 
in Table~\ref{tb:B.E.}
the calculated binding energies and level widths
for $0s$ and $0p$ states 
in the
$^{11}$B and $^{39}$K cases. We also show 
in Table~\ref{tb:B.E.} the binding energies and widths by
the chiral unitary approach reported in Ref.~\cite{PLB550} 
for comparison.
The level 
structure of 
the bound states is 
qualitatively 
very similar with
that of 
the chiral doublet model with $C=0.0$
as we can expect from the potential shape. For the $C=0.2$ case
in the chiral doublet model, 
we can not find 
any
bound states because of the repulsive nature of the optical potential.

\begin{table}
\begin{tabular}[t]{ccccc}
\multicolumn{3}{l}
{Chiral doublet model with C=0.0 [MeV]}\\
\hline
\hline
& $^{11}$B & $^{39}$K \\
\hline
$0s$ & (13.7, 41.5)  & (30.3, 42.5) \\
$0p$ & -  & (14.6, 50.7) \\
\hline
\end{tabular}
\begin{tabular}[t]{ccccc}
\multicolumn{3}{l}
{Chiral doublet model with C=0.2}\\
\hline
\hline
& $^{11}$B & $^{39}$K \\
\hline
\multicolumn{3}{l}
{\it no bound state}\\
\hline
\end{tabular}
\\\\\\
\begin{tabular}[t]{ccccc}
\multicolumn{3}{l}
{Chiral unitary approach \cite{PLB550} [MeV]}\\
\hline
\hline
& $^{12}$C & $^{40}$Ca \\
\hline
$0s$ & (9.71, 35.0)  & (17.9, 34.4) \\
$0p$ & -  & (7.0, 38.6) \\
\hline
\end{tabular}
\caption{
Binding energies and widths calculated by the Chiral
 Doublet model (C=0.0 and C=0.2)
with the mirror assignment \cite{PRC66(02)045202}
and by the chiral unitary
 approach reported in Ref.~\cite{PLB550}.
\label{tb:B.E.}
}
\end{table}

\subsection{$\omega$-nucleus optical potential}

Let us move to 
discussion on the optical potential of 
in-medium $\omega$ meson. In 
our 
study of the $\omega$ mesic nuclei, we use a
simple model for the optical potential, which was used in the previous study
\cite{EPJA6} on the $\omega$-mesic nuclei formation using the (d,$^3$He)
reaction. 
An
empirical potential is given in an energy-independent
$T\rho$ approximation as
\begin{equation}
V_\omega^{\rm (a)}(r)=-(100+70i)\frac{\rho(r)}{\rho_0}. \label{eq:OPomega}
\end{equation}
The real and imaginary parts of the optical potential were estimated based on
15 \% mass reduction of the $\omega$ mass
at the normal density and the $\omega$ life
time ($\tau=1.5$ fm/c) of nuclear absorption in the nuclear medium,
respectively, as reported in Ref.\cite{NPA624}. 

The binding energies and widths of the omega meson in $^{11}$B
are
calculated by solving the Klein-Gordon equation~(\ref{eq:K-Geq}) with the above potential.
In this potential, we find two bound states in the $^{11}$B nucleus. 
The obtained binding energies and widths are shown in Table~\ref{tab:omegaBE}. 
Since the width of these states are larger than the level spacing,
it would be difficult to observe distinct peak structure in
inclusive spectra.
%
In next section, we 
will
show the calculated ($\gamma$,p) spectra for several incident energies. 

\begin{table}
\begin{tabular}{ccc}
\hline \hline
$\omega$-$^{11}$B bound states
   & $V_\omega^{\rm (a)}$ & $V_\omega^{\rm (b)}$\\
\hline
$0s$ &    (49.0, 116) &  (103, 54.4)\\
$1s$ &    -    &     (18.8, 27.6) \\
$0p$ & (14.3, 86.5) &  (58.1, 42.7)\\
$0d$ &  - &   (17.0, 30.5)\\
\hline \hline
\end{tabular}
\caption{Binding energies and widths of the $\omega$ bound states in 
$^{11}$B with
the
optical potentials (\ref{eq:OPomega}) and (\ref{eq:OPomega2}) in unit of
MeV.
\label{tab:omegaBE}}
\end{table}

We also consider two more optical potentials in order to investigate the
sensitivity of the depth and the absorption of the potential to the ($\gamma$,p)
spectra. The potentials considered here are 
\begin{eqnarray}
  V_\omega^{\rm (b)}(r)&=&-(156+29i)\frac{\rho(r)}{\rho_0}\ , \label{eq:OPomega2}\\
   V_\omega^{\rm (c)}(r)&=&-(-42.8+19.5i)\frac{\rho(r)}{\rho_0} \label{eq:OPomega3}\
\end{eqnarray}
which are obtained by the linear density approximation with the scattering
lengths $a=1.6+0.3i$ fm \cite{NPA650} and $a= -0.44 + 0.2i$ fm
\cite{Lutz:2001mi}, respectively.
The former scattering length is obtained by the $\omega N$ scattering
amplitude calculated at one loop approximation with an effective Lagrangian
based on the SU(3) chiral symmetry incorporating vector mesons. On the other
hand, the later is obtained in a relativistic and unitary approach for meson-baryon amplitudes.
In this case the scattering length is to be repulsive 
due to a subthreshold effect of the $\omega N(1520)$. 

The energies and widths of 
the bound states calculated with the potential $V_\omega^{\rm (b)}$ are
also shown 
in Table~\ref{tab:omegaBE}. In this case, more bound states are found with relatively narrower
widths than the case of the previous potential $V_{\omega}^{\rm (a)}$
since the optical potential $V_\omega^{\rm (b)}$ has deeper real
potential and less nuclear absorption.   
The magnitudes of the level widths are comparable to those
of level spacing and, hence, we expect to observe peak structure in
reaction spectra.
Later we shall compare the
($\gamma$,p) spectra calculated with these potentials at the incident energy
$E_{\gamma} = 2.7$ GeV where the recoilless condition is satisfied.

%

\subsection{$\sigma$-nucleus optical potential}
\label{sec:sigma}
Finally, we consider the $\sigma$ meson-nucleus systems.
%
%
In this paper we take a particle picture of the sigma meson, in which the
sigma meson is described as a particle with a huge decay width to two pions.
The calculation of the self-energy of the sigma meson is performed based on the
SU(2) linear sigma model \cite{PRL82},
and the partial restoration of chiral symmetry in nuclear medium
is parameterized as the sigma condensation in the medium within the model.  

The optical potential for the $\sigma$-nucleus is defined
as the density-dependent part of the in-medium $\sigma$ self-energy
divided by 
two times
the bare sigma mass $m_\sigma$ as;
\begin{equation}
\label{eq:sigopt}
{\rm Re} V_{\sigma}(\omega, \rho)  = 
      - \frac{1}{2m_\sigma} \lambda \sigma_0^2 \ 
\left( 1 - \Phi(\rho) \right).
      \label{eq:RealOptSig}
\end{equation}
Here we use the $\sigma$ self-energy evaluated in Ref.~\cite{PRL82} 
based on the SU(2) linear sigma model at the one-loop level,
which corresponds to the mean field approximation.
The density dependent function $\Phi(\rho)$ is defined to express the chiral
condensate in nuclear matter $\langle\sigma\rangle_{\rho}$ as
$\langle\sigma\rangle_{\rho} = \sigma_0 \Phi (\rho)$.
The function $\Phi(\rho)$
is given in Eq.(\ref{eq:defPhi}) with the $C$ paramater,
which represents the strength of the partial restoration of the chiral
symmetry in the nuclear medium.
It is worth noting that at the one-loop approximation the imaginary part of
the self-energy is independent of the density, so that the nuclear absorption of
the sigma meson appears from the next order, 
such that multi-{\it ph} excitations and
pion - {\it ph} excitations.

The density-independent part of the self-energy $\Sigma_{\sigma}$ is
given in the one-loop calculation in the linear sigma model as 
\begin{eqnarray}
{\rm Re} \Sigma_{\sigma}(\omega)  = 
   &-&   {\lambda \over 32 \pi^2} \  \left[ 
               m_{\pi}^2 (1- \ln {m_{\pi}^2 \over \kappa^2})
              +m_{\sigma}^2 \ (1- \ln {m_{\sigma}^2 \over \kappa^2})
         \nonumber \right. \\
 &  +  &  {1 \over 3} \lambda \sigma_0^2 \  (Q_{\pi} +
              2- \ln {m_{\pi}^2 \over \kappa^2}) \nonumber \\
  &  +  & \left. \lambda \sigma_0^2 \ (Q_{\sigma} +
              2- \ln {m_{\sigma}^2 \over \kappa^2}) \right] , 
\label{eq:Sigma_sigma}\\
{\rm Im} \Sigma_{\sigma}(\omega)  = 
  &-& {\lambda^2 \over 32 \pi}  \sigma_0^2 \
  \left[
    {1  \over 3}   P_{\pi} \theta (\omega - 2 m_{\pi} )
    +  P_{\sigma} \theta (\omega - 2 m_{\sigma} ) \right] ,
\label{eq:SigmasigmaIm}
\end{eqnarray}
where 
the function
$Q_{\varphi}$ is given as  
\begin{equation}
Q_{\varphi}  =  
\left\{
   \begin{array}{l}
   P_{\varphi} \ \ln \ \displaystyle\frac{1-P_{\varphi}}{1+ P_{\varphi}} \ \ \ 
         ({\rm for} \ \ 2m_{\varphi} \le \omega ) \\
   -2 P_{\varphi} \ {\rm arctan} \ \displaystyle\frac{1}{P_{\varphi}} \ \ \ 
         ({\rm for} \ \   \omega \le 2m_{\varphi}  ) 
   \end{array}
\right.,
\label{eq:Q_phi}
\end{equation}
with $P_{\varphi}  =  \left| 1 - 4 m_{\varphi}^2 / \omega^2 \right| ^{1/2}$.
 Here,  $\varphi$ denotes either $\pi$ or $\sigma$,
 $\kappa$ is a renormalization point
 in the minimal subtraction scheme and 
$\lambda$ is the coupling constant of 
the
four-point meson vertex in the
linear sigma model.
The reduced mass $\mu$ shown in the Klein-Gordon equation (\ref{eq:K-Geq}) is
corresponding to $\mu^{2} = m_{\sigma}^{2} + {\rm Re}\Sigma_{\sigma}$ 
in heavy nucleus limit
for
the sigma meson case. 
The bare sigma mass $m_{\sigma}$ is fitted as 469 MeV so that the 
sigma meson mass in free space $m_\sigma^{\rm (free)}$ is reproduced as
550 MeV after the one-loop calculation. 
%
All parameters appeared in Eqs.~(\ref{eq:Sigma_sigma})-(\ref{eq:Q_phi}) are
obtained in the first reference of Ref.\cite{PRD57etc} and
listed in Table.1 of
Ref.\cite{NPA710}.

\begin{figure}[hbt]
\epsfysize=5cm
\centerline{
\epsfbox{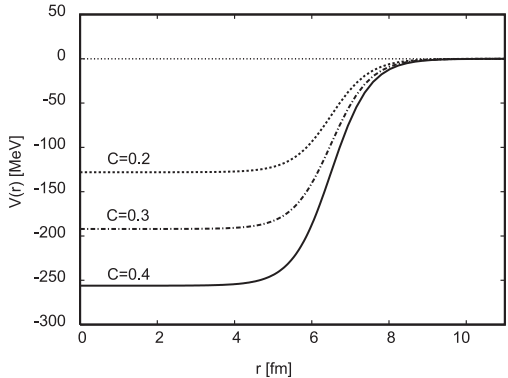}
}
\caption{
Real potential for the $\sigma$ meson inside the $^{208}$Pb nucleus
 as a function of the radial coordinate $r$ obtained in Ref.~\cite{NPA710}. 
The density is assumed to be the Woods-Saxon form with $R=6.5$ fm and
 $a=0.5$ fm in Eq.~(\ref{eq:W-S}).
Each line indicates $C=0.2$ (dotted line), $0.3$ (dot-dashed line) and
 $0.4$ (solid line) case, respectively.
}
\label{fig:sigma_pot}
\end{figure}

The self-energy of the sigma meson obtained in the mean field provides a strong attractive potential for the sigma meson in medium. 
We show in Fig.~\ref{fig:sigma_pot} the attractive potential
Re$V_{\sigma}(r)$, given in Eq.(\ref{eq:RealOptSig}),
for $C=0.2$, $0.3$ and $0.4$ in $^{208}$Pb.
We assume that 
the density distribution is 
of the Woods-Saxon type as defined in
Eq.~(\ref{eq:W-S}) with $R=6.5$ fm and $a=0.5$ fm.  
It is seen that the depth of the attractive potential is strongly depend on the $C$ parameter.
The strong attraction of the optical potential in the center of nucleus appears
also in lighter nuclei.

As a result of the attraction, we expect that a strong reduction of the $\sigma$ mass 
approaching down to the two pion threshold especially for heavy nuclei. Consequently there is a chance to form sigma bound states with narrower widths. 
The reduction of the sigma mass in nuclear medium is
expected also in the $\pi\pi$ resonance picture of the sigma meson with the
non-linear realization of chiral symmetry, if one takes account of appropriate
medium effects through the pion wave function renormalization \cite{PRD63}.

\begin{figure}[hbt]
\epsfxsize=12cm
\centerline{
\epsfbox{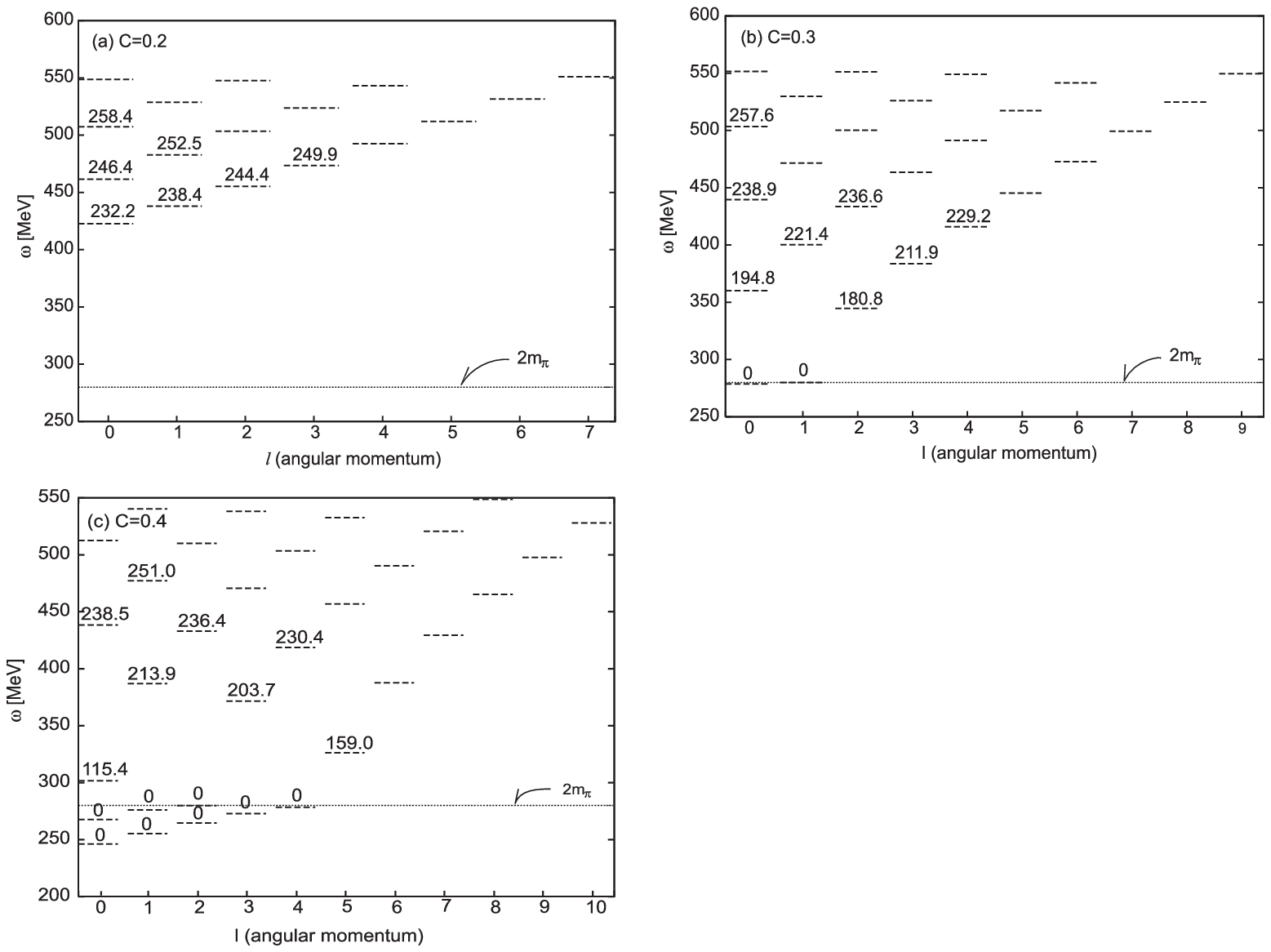}
}
\caption{Calculated eigenenergies of the bound $\sigma$ in $^{208}$Pb~\cite{NPA710}
 with (a) $C=0.2$, (b) $C=0.3$ and (c) $C=0.4$. The $\ell$ is the orbital angular
 momentum of the $\sigma$ states. Widths are shown by numbers for low lying
 states in unit of MeV. The $2m_\pi$ threshold is shown by the dotted
 line. The states blow the threshold do not have widths due to the
 $\pi\pi$ decay in the present calculation.}
\label{fig:sigma_B.E.}
\end{figure}

The energies and widths of the sigma bound states are calculated
by solving the Klein-Gordon equation for the $\sigma$ meson with
the optical potential $V_{\sigma}$ and the self-energy $\Sigma_\sigma$ with the one-loop correction.
%
In Fig.~\ref{fig:sigma_B.E.}, we show the bound state spectra of the
$\sigma$ meson in $^{208}$Pb
with $C=0.2$, $0.3$ and $0.4$ cases, together with widths of low
lying states~\cite{NPA710}.
It is found that, if the partial restoration of 
chiral symmetry 
in nuclear medium
occurs with sufficient strength, deeply bound $\sigma$ states in heavy nuclei 
formed with 
significantly smaller decay widths than 
that
in vacuum.
%
This is responsible for
suppression of the available $\pi\pi$ phase space associated with the
reductions of the real energies of the bound states.
Furthermore, for $C=0.3$ and $0.4$ cases, there are several bound states 
with no decay widths
below
the $2\pi$ threshold 
since the $\sigma\rightarrow\pi\pi$ decay channel, which is the only possible
decay mode at low energies in the present model, is kinematically forbidden. 
%

In order to see a global structure of the $\sigma$ bound states in the nucleus and the impact of the narrow bound states on the production reaction, we show the spectral functions of the bound states, which are used to calculate the $\sigma$ production spectra in the ($\gamma$,p) reaction later. The spectral function of the $(n,\ell)$ bound state is given by
\begin{equation}
\label{eq:spe_func}
   \rho_{n\ell}(\omega)  = - \frac{1}{\pi} \frac{{\rm Im} \Sigma_{\sigma}(\omega)}
   {(\omega^{2} - \omega^{2}_{n\ell})^{2} + 
({\rm Im} \Sigma_{\sigma}(\omega))^{2}} \ ,
\end{equation}
where $\omega_{n\ell}$ denotes the eigenenergy of the $\sigma$ bound
state in the nucleus.  The spectral functions Eq.~(\ref{eq:spe_func}) encounter a divergence at $\omega=\omega_{n\ell}$ for eigenstates in case of Im$\Sigma_\sigma=0$. This is not the case if we take into account nuclear absorptions of the sigma meson which we do not consider here. In order to regularize the divergence, we add
an extra 5 MeV width to the Im$\Sigma_\sigma$ in all the bound states in the following calculations. 

\begin{figure}[hbt]
\epsfxsize=7cm
\centerline{
\epsfbox{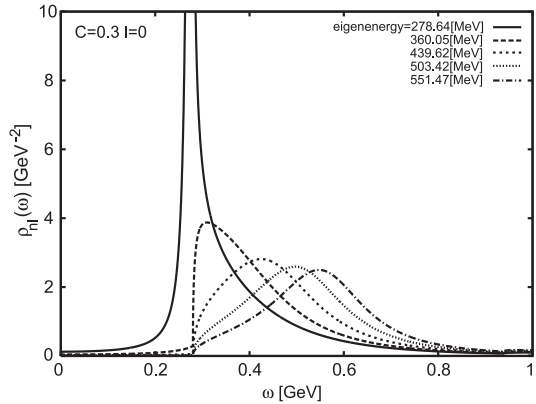}
}
\caption{
Spectral functions $\rho_{n\ell}(\omega)$ of the $\sigma$ bound
 states with $\ell=0$ and $n=1,2,\dots,5$ for $C=0.3$ case obtained in
 Ref.~\cite{NPA710}. An extra 5 MeV width is added to the imaginary 
 part of the self-energy.
}
\label{fig:spe_func}
\end{figure}

Figure~\ref{fig:spe_func} shows a series of $\rho_{n\ell}$ for the $\ell=0$ bound states in $^{208}$Pb for the $C=0.3$ case~\cite{NPA710}. As seen in the figure, 
the spectral functions for the deeply bound states have a prominent peak around $\omega\sim 2m_\pi$. This is a direct consequence that the deep bound states are formed as a result of the strong attractive potential.
The detail discussions are given in Ref.~\cite{NPA710}. The appearance
of the narrow states at the low energies gives also strong enhancement
of the sigma production rate in nucleus around the two pion threshold,
which is expected to be seen in the ($\gamma$,p) reaction. This point is
discussed in the following section.  

It is also possible to form the deeply bound states with the narrow width
in lighter nuclei than Pb, since the potential depth is nearly independent of the
mass number of nucleus for a certain value of the $C$ parameter.
The spatial range of the potential,
however, depends on nuclear size.
Therefore,
the zero width bound states
are expected to be
formed in nuclei
which are larger
than a certain spatial dimension.

\section{Formation of meson-nucleus system in ($\gamma$,p) reaction}
\label{sec3}
In this section we explain the advantages of the ($\gamma$,p) reaction for investigation of the meson-nucleus system and briefly review the formulation to obtain the formation rate of the mesons inside nucleus.
\subsection{($\gamma$,p) reaction}

\begin{figure}[hbt]
\epsfysize=5cm
\centerline{
\epsfbox{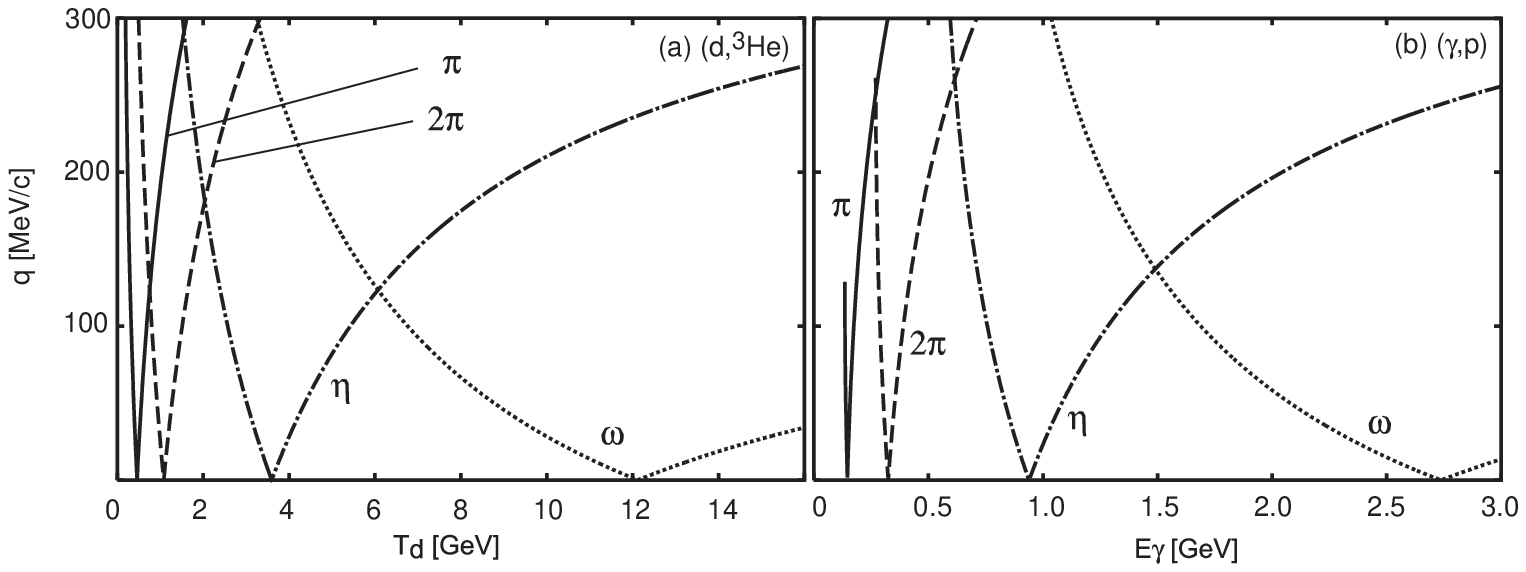}
}
\caption{
Momentum transfer $q$ is shown as a function of (a) the incident deuteron
 kinetic energy 
 $T_d$ in the (d,$^3$He) reaction, and (b) the incident $\gamma$ energy
 $E_\gamma$ in the ($\gamma$,p) reaction.
The solid line shows the momentum transfer of the pion production case,
and other lines show those of two pions, the $\eta$ meson and the
 $\omega$ meson production cases as indicated in the figure.
}
\label{fig:Q}
\end{figure}
In 
the ($\gamma$,p) reaction,
the incident photon reacts with a proton inside the
target nucleus,
producing a meson 
there most likely at rest with recoilless condition.
The meson may be
trapped
%
into a certain bound state in the nucleus, if there exist bound states.
The proton is ejected from the nucleus and 
is to be
observed at a forward angle.
In the ($\gamma$,p) reaction spectroscopies, 
one observes only
the emitted proton in the final state 
for simplicity of the experiment
and obtains the double differential
cross section $d\sigma/d\Omega/dE$ as a function of the emitted proton
energy. 
Investigating the differential cross section, we 
extract the
properties of
the interaction between each meson and the nucleus.
Observing the more particles in the final state, we will obtain
the more information of the meson-nucleus interactions. 

The incident energy for each meson production is tuned so as to satisfy the
recoilless condition achieved by the zero momentum transfer, in which the
meson may be produced in the nucleus at rest. 
In the recoilless condition, due to 
the matching condition between momentum- and angular momentum-transfer,
we have
a selection rule for the total angular momentum of the meson and the proton hole
states. Namely the contributions of the different angular momenta 
are strongly suppressed. 

In Fig.~\ref{fig:Q}, we show the momentum transfers as
functions of the incident particle energies for the (d,$^3$He) and
($\gamma$,p) reactions. In the plots we use the meson masses in free space.
In the following discussion for the ($\gamma,p$) reaction we choose the incident
energy $E_{\gamma}=950$ MeV for the eta meson production and 
$E_{\gamma}=2.75$ GeV for the omega meson production to satisfy the
recoilless condition for the corresponding meson-nucleus systems. For the sigma meson case, the incident 
energy is chosen to be $E_{\gamma}=400$ MeV, which is a recoilless condition
for the two pion production in the nucleus, since we focus our interest on the spectral
function enhancement near the two pion threshold in the sigma channel 
associated with the partial restoration of chiral symmetry in nuclear medium.

%

\begin{figure}[hbt]
\epsfxsize=5cm
\centerline{
\epsfbox{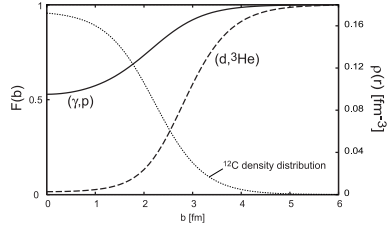}
}
\caption{Distortion factor ${\bar F}(b)$
for the $\eta$ meson production reactions in $^{12}$C target
as a function of the impact
parameter $b$. The solid line indicates the ${\bar F}(b)$ for the ($\gamma$,p)
reaction and the dashed line for the (d,$^3$He) reaction.
The density profile of $^{12}$C is also shown.
}
\label{fig:distortion}
\end{figure}

The major advantage of use of the ($\gamma$,p) reaction for the mesic nucleus
formation is the smaller distortion effect on the incident photon than
hadronic beams. The distortion effects of the projectile and ejectile are
known
to reduce significantly the reaction rate inside the nucleus.
As we shall discuss later in Fig.\ref{fig:distortion}, half
of the 
flux
of the ($\gamma$,p) reaction
reaches the center of the target nucleus, while most of the
flux in the (d,$^{3}$He) reaction is
restricted
on the surface of the
nucleus. Therefore, due to the transparency of photon, we can probe the 
meson wavefunction in the
center of the nucleus with the ($\gamma$,p) reaction. This characteristic
feature is
desirable
especially to the study of the eta-nucleus system, in
which we have the discrepancy in the model calculations of the $\eta$ optical
potential at the center of the nucleus depending on the physical picture of
$N^{*}$, as discussed in Sec.\ref{sec:etapot}

The other advantage of the ($\gamma$,p) reaction is that, due to the transparency
of photon, we expect more production rate of the low lying states of the mesic
nucleus. This is favorable to the sigma-nucleus system, since we expect the
deeply bound states of the sigma meson with a significantly narrow width as a result of strong
reduction of the chiral condensate in the nucleus. 
In the case of mesic nuclei formation, the meson wave function in nucleus
resembles the proton hole wave function
in the sense of function orthogonality,  
because both of the wave functions are,
roughly to say,
solutions of the Schr\"odinger equations with the same square-well type
potentials. (Recall that the potentials for the mesons is obtained as the 
Woods-Saxon type shape under the local density assumption with the
nuclear density distribution (\ref{eq:W-S}) except the case of the eta meson
with the chiral doublet model $C=0.2$.) Due to the orthogonality of the
wave functions, we have the selection rule that the combinations of the
different
principle
quantum numbers of the meson and proton hole wave functions are strongly
suppressed in the recoilless condition~\cite{EPJA6,NPA710}. 
This is reminiscent of the recoilless formation of substitutional states in
hypernuclear reactions. Therefore, in order to produce the ground state of the
mesic nucleus, in which the meson sits on the $s$ state, we need to pick up
the deepest bound proton from the target nuclei using a better transparent
incident beam. 
In contrast, in the case of the atomic state formation, 
the orthogonal condition of the radial wave functions of the meson and nucleon
is not 
satisfied
any more, since the meson is trapped in a
Coulomb potential and the meson distribution is roughly expressed by the Coulomb
wave function.
Then the recoilless condition enhances the population of any states
which have the total angular momentum $J\sim 0$, for example, in the
formation of the deeply bound pionic atoms in Sn
isotopes~\cite{PLB514,PTP103}, the contribution of $(3s)_n^{-1}\otimes
1s_\pi$ combination can be largely populated in the (d,$^3$He) reaction.

Let us evaluate the distortion effects of the (d,$^3$He) and
($\gamma$,p) reactions to compare the 'transparency' of both reactions.
%
We describe the distorted waves of the incoming projectile and of the
outgoing ejectile as $\chi_{i}$ and $\chi_{f}$, respectively. Using the Eikonal
approximation, we write
\begin{equation}
\chi_f^*({\bf r})\chi_i({\bf r})=\exp(i{\bf q}\cdot{\bf r})F({\bf r}),
\label{eq:eikonal}
\end{equation}
with the momentum transfer $\bf q$. 
We choose the beam direction as to be $z$-axis. The momentum transfer
{\bf q} is along $z$-axis in the forward reactions which we consider
here. 
The distortion factor $F({\bf r})$
%
%
 is defined as,
\begin{equation}
F({\bf r})=\exp\left[-\frac{1}{2}\sigma_{iN} \int_{-\infty}^z dz' \rho_A(z',b)
-\frac{1}{2}\sigma_{fN} \int^{\infty}_z dz' \rho_{A-1}(z',b)
\right].
\label{eq:distfac}
\end{equation}
where $\sigma_{iN}$ ($\sigma_{fN}$) is the total cross section of the
nucleon and incident (emitted) particle and $\rho_{A}(z,b)$ is the
density distribution function 
for the nucleus with the mass number $A$
in the coordinates of $z=r \cos\theta$ and $b=r\sin\theta$.
The distortion factor, which will be used in the calculation of the
($\gamma$,p) reaction cross section later, gives an estimation 
of the reduction of the beam flux 
caused by
the nuclear absorption
effects. In order to take a view of the distortion effect, we calculate
an averaged distortion factor, which is defined as
\begin{equation}
\bar F(b)=\exp\left[
-\frac{1}{2}\bar{\sigma}\int_{-\infty}^\infty\bar{\rho}(z',b) dz'
\right],
\end{equation}
where
$\bar{\sigma}$ is an average distortion cross section
of the initial and final channels, and
$\bar{\rho}(z,b)$ denotes an average nuclear density of the target and
daughter nuclei at 
an impact parameter $b$ and a beam direction coordinate $z$.

In Fig.~\ref{fig:distortion}, we show the averaged distortion factors $\bar F(b)$
in 
the ($\gamma$,p) reaction and the (d,$^3$He) reaction 
of
the $^{12}$C target case for $\eta$ production,
which
will be 
discussed in Sec.~\ref{sec:eta}.
Here we have used $\bar \sigma =104$ mb for the (d,$^{3}$He) reaction and
$\bar \sigma = 16$ mb for the ($\gamma$,p) reaction. 
These cross sections are evaluated from experimental
nucleon-nucleon scattering
data at appropriate energies~\cite{ParticleData}.
%
The figure shows that 
$\bar F(b)$ in the ($\gamma$,p) reaction has a finite value ($\geq 0.5$) even in the
nuclear center. This means that the amplitude of the meson production
at center of nucleus is suppressed only by half due to the nuclear
absorption. On the other hand, the averaged distortion factor of the
(d,$^3$He) reaction is almost zero inside the 
nucleus. 
Therefore
the photon can reach the center of the nucleus and
create the meson there, while the deuteron mostly creates meson only on the surface of
the nucleus. 
Thus, as we expected,  the ($\gamma$,p) 
spectra are shown to be more sensitive to 
the optical potential of the meson at
the interior of the nucleus. 
%

\subsection{Formulation}
\label{sec:formulation}
We use the Green function method\cite{NPA435} to calculate the formation cross
sections of the $\eta$-nucleus and the $\omega$-nucleus systems in the
($\gamma$,p) reaction. The details of the application of the Green function
method for the (d,$^{3}$He) reaction are found in
Refs.~\cite{PRC66(02)045202,EPJA6,NPA650}. 

The present method starts with a
separation of the reaction cross section into the nuclear response function
$S(E)$ and the elementary cross section of the $p$($\gamma$,p)$\eta$ or
$\omega$ with the impulse approximation:
\begin{equation}
  \left( \frac{d^{2}\sigma}{d\Omega dE}\right)_{A(\gamma,p)(A-1)\otimes \varphi} 
  =\left(\frac{d\sigma}{d\Omega}\right)_{p(\gamma,p)\varphi}^{\rm lab}  \times
S(E) \ , \label{eq:Green}
\end{equation}
where $\varphi$ denotes the $\eta$ or $\omega$ meson. 
The calculation of the nuclear response function with the complex potential is
formulated by Morimatsu and Yazaki \cite{NPA435}
in a generic form as
\begin{equation}
  S(E) = -\frac{1}{\pi} {\rm Im} \sum_{f} {\cal T}_f^{\dagger} G(E) {\cal T}_f
\end{equation}
where the summation is taken over all possible final states. The
amplitude ${\cal T}_f$ denotes the transition of the incident particle to
the proton hole and the outgoing ejectile, involving the proton hole
wavefunction $\psi_{j_p}$ and the distorted waves, $\chi_{i}$ and
$\chi_{j}$, of the projectile and ejectile, taking the appropriate spin
sum: 
\begin{equation}
  {\cal T}_f({\bf r}) = \chi^{*}_{f}({\bf r}) \xi^{*}_{1/2,m_{s}} 
  \left[Y_{l_{\varphi}}^{*}(\hat r)\otimes \psi_{j_{p}}({\bf r})\right]_{JM} \chi_{i}({\bf r})
\end{equation} 
with the meson angular wavefunction $Y_{l_{\varphi}}(\hat r)$ and the
spin wavefunction $\xi_{1/2,m_s}$ of the ejectile.
The distorted waves are written with the distortion factor as
Eq.(\ref{eq:eikonal}) in the Eikonal approximation. The Green function
$G(E)$ contains the meson-nucleus optical potential in the Hamiltonian
as 
\begin{equation}
   G(E; {\bf r}, {\bf r^{\prime}}) = \langle p^{-1} | \phi_{\varphi}({\bf r})
\frac{1}{E-H_{\varphi}+i\epsilon} \phi_{\varphi}^{\dagger}({\bf r^{\prime}}) |
p^{-1} \rangle
\label{eq:Green_function}
\end{equation}
where $\phi^{\dagger}_{\varphi}$ is the meson creation operator and
$|p^{-1}\rangle$ is the proton hole state. Obtaining the Green function
with the optical potential is essentially same as solving the associated
Klein-Gordon equation.   
We can calculate the nuclear response function $S(E)$ from ${\cal
T}_f^\dag({\bf r}) G(E;{\bf r},{\bf r'}){\cal T}_f({\bf r'})$ by performing
appropriate numerical integrations for variables ${\bf r}$ and ${\bf r'}$.
%
%

%
For the $\sigma$-nucleus systems, we do not use the Green function
method to evaluate the formation rate, because (a) we do not know the
elementary cross section of the $\gamma + p \rightarrow \sigma + p$
reaction yet and (b) the sigma optical potential has a large imaginary
part even outside of the nucleus due to the large decay width, of which
boundary condition the Green function does not work in the present
formalism. Alternatively   
%
we evaluate the spectral function of the $\sigma$ bound states defined as
\begin{equation}
  \rho_{\sigma}(E) =  \sum_{n,\ell} (2l+1) \rho_{n \ell}(E) \ ,
\end{equation}
where $\rho_{n\ell}(E)$ denotes the contribution from each bound state
in the nucleus with the $(n,\ell)$ quantum number as given in
Eq.~(\ref{eq:spe_func}). 
In order to take  the reaction mechanism into consideration, we
introduce an effective number $N_{\rm eff}$, which represents relative
production wight of each bound state \cite{NPA710,NPA530etc}.  
Finally we obtained the total spectral function as
\begin{equation}
\rho_{\rm tot}(\omega)=\sum_{n\ell} N_{\rm eff}\rho_{n\ell}(\omega).
\label{eq:tot_spe}
\end{equation}
with the effective number $N_{\rm eff}$ given by
\begin{equation}
\label{eq:effNum}
N_{\rm eff}=\sum_{JMm_s}|\int d^3r
\chi^*_f({\bf r})\xi_{1/2,m_s}^*
\left[\phi^*_{\ell_\sigma}({\bf r})\otimes\psi_{j_n}({\bf r})
\right]_{JM}
\chi_i({\bf r})|^2
\end{equation}
where $\psi_{j_p}$ denotes the wave function of the picked-up proton and,
$\phi_{\ell_\sigma}$ and $\xi_{1/2,m_s}$ are the $\sigma$ wave function
and the spin wave function of the ejectile, respectively. The $\sigma$
wave function is obtained by solving the Klein-Gordon equation with the
optical potential and the self-energy shown in Eqs.~(\ref{eq:sigopt}),
(\ref{eq:Sigma_sigma}) and (\ref{eq:SigmasigmaIm}). 

This effective number
approach is 
known to be
a good approximation
of the Green function method
to calculate
the reaction cross section in Eq.(\ref{eq:Green}),
if the bound states lie well-separately with narrow
widths~\cite{NPA435}, and applied to evaluate the formation cross
sections of deeply bound pionic atoms~\cite{NPA530etc}. 
In the present case of the
sigma meson in nucleus, we expect the 
narrow bound states around the two-pion threshold for the $C=0.3$ and
$C=0.4$ cases, and focus our interest on seeing the impact of the narrow
bound states on the ($\gamma$,p) reaction. Therefore 
the approach with the
spectral function is expected to be reasonable for our purpose.


%
\section{Result}
\label{sec4}
In this section, we evaluate the formation rates of the
mesic nuclei
by the ($\gamma$,p) reactions and show the 
spectra calculated with the optical potentials discussed in Section \ref{sec2}
for the $\eta$-, $\omega$-, and $\sigma$-nucleus systems.

\subsection{$\eta$-nucleus system}
\label{sec:eta}
We discuss in this section the formation rate of the $\eta$-mesic
nucleus in the ($\gamma,p$) reaction.  
The initial photon energy is chosen as the recoilless condition energy
$E_\gamma=950$ MeV 
as shown in Fig.~\ref{fig:Q}(b).
The emitted proton spectra around the $\eta$ production energies
are evaluated in the Green function method (\ref{eq:Green})
with the $\eta$
Green function (\ref{eq:Green_function})
calculated with the optical
potential~(\ref{eq:eta-potential}).
%
We estimate the elementary cross section $d\sigma/d\Omega$ of the
$\gamma +p\rightarrow p+\eta$ reaction at this energy in the laboratory
frame as to be 
$3.4 \mu$b/sr using the experimental data by the tagged photon beam at
University of Tokyo~\cite{JPSJ57}.

The obtained spectra are shown as functions of the excitation energy
$E_{\rm ex}$
measured from the eta meson production threshold energy $E_{0}$. The excitation energy
corresponds to
the missing energy in the inclusive reaction $\gamma + A \rightarrow p
+ X$. In the plot of the separated contributions of the $\eta$-hole
configurations,  it should be taken into account the appropriate
hole energy depending on the orbit $j_{p}$ in the daughter nucleus. 
The hole energy $S_p(j_p)-S_p({\rm ground})$ is measured from the ground state
of the daughter nucleus, which are listed in Table \ref{tab:sepene}. The binding
energy $B_\eta$ of the $\eta$ meson is deduced from the excitation energy and
the hole energy as
\begin{equation}
E_{\rm ex}=m_\eta-B_\eta + [S_p(j_p)-S_p({\rm ground})].
\label{eq:Eex-E0}
\end{equation}

\begin{table}
\begin{tabular}{|c|c|c|c|}
\hline\hline
$^{12}$C & $j_p$ & $S_p(j_p)-S_p(1p_{3/2})$ & $\Gamma_{j_p}$ \\
\hline
 & $1p_{3/2}$ &  0 & -  \\
 & $1s_{1/2}$ & 18 & -  \\
\hline\hline
$^{40}$Ca & $j_p$ & $S_p(j_p)-S_p(1d_{3/2})$ & $\Gamma_{j_p}$ \\
\hline
 & $1d_{3/2}$ & 0 & 0 \\
 & $2s_{1/2}$ & 3.2 & 7.7 \\
 & $1d_{5/2}$ & 8.0 & 3.7 \\
 & $1p_{1/2,\ 3/2}$ & 24.9 & 21.6 \\
 & $1s_{1/2}$ & 48.0 & 30.6 \\
\hline
\end{tabular}
\caption{Proton separation energies of the single particle orbits 
measured from the ground state of the
daughter nucleus are shown in unit of MeV.
Data taken from Ref.~\cite{12C_ref} for $^{12}$C and Ref.~\cite{40Ca_ref} for
 $^{40}$Ca. Widths of the proton hole states deduced from the data are
 also shown for $^{40}$Ca.
\label{tab:sepene}}
\end{table}

\begin{figure}[hbt]
\epsfxsize=12cm
\centerline{
\epsfbox{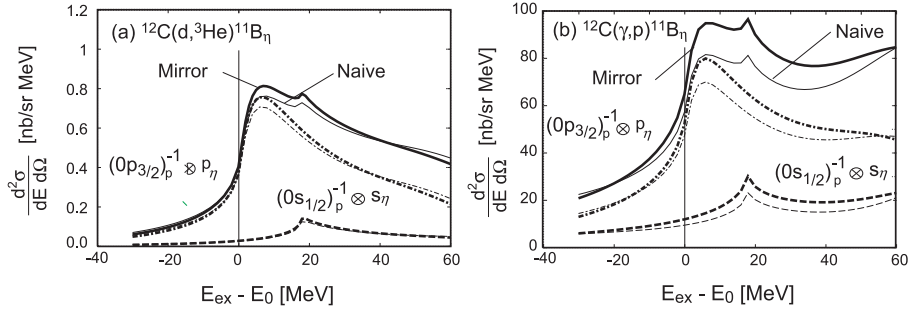}
}
\caption{
Calculated $\eta$-production spectra of the (a) (d,$^3$He) reaction
 at $T_d=3.5$ GeV and (b) ($\gamma$,p) reaction at $E_\gamma=950$ MeV
 for the $^{12}$C target are shown as
functions of the excitation energy
$E_{\rm ex}$ defined in the text.
$E_0$ is the $\eta$ production threshold energy. The $\eta$-nucleus
 interaction are evaluated using the chiral doublet model with the
 mirror assignment (thick lines) and naive assignment (thin lines) with $C=0.2$. 
The total spectra are shown by the solid lines and the dominant
 contributions from the $(0s_{1/2})_p^{-1}\otimes s_\eta$ and
 $(0p_{3/2})_p^{-1}\otimes p_\eta$ configurations are shown by the
 dashed lines and dash-dotted lines, respectively, where the proton-hole
 states are indicated as $(n\ell_j)_p^{-1}$ and $\eta$ states as $\ell_\eta$.
}
\label{fig:MvsN}
\end{figure}

We evaluate the differential cross section of the ($\gamma$,p) reaction on the
$^{12}$C target.
The $\eta$ meson is created in the $^{11}$B nucleus
because the proton is knocked out in the reaction. 
In Fig.~\ref{fig:MvsN}, we show the results calculated by the chiral doublet
model with both mirror and naive assignments  together
with those of the (d,$^3$He) reaction 
reported in 
\cite{PRC66(02)045202,PRC68(03)035205} 
for $C=0.2$ case in Eq.(\ref{eq:defPhi}).
%
%

First of all we would like to mention that the peak structure seen in the spectra has nothing to do with the existence of the $\eta$ bound states. (Actually in the case of $C=0.2$ there are no bound states.) These peaks are just signals of opening the phase space of the $\eta$ creation in the nucleus \cite{PRC68(03)035205}. 

Secondly we see
in 
the figure that
each spectrum is dominated by two
contributions, $(0s_{1/2})_p^{-1}\otimes s_\eta$ and
$(0p_{3/2})_p^{-1}\otimes p_\eta$
%
This is a consequence 
of the
matching condition
in the recoilless
kinematics
that 
final states with total spin $J\ne 0$ are largely suppressed.
The $\eta$ production threshold for the $(0p_{3/2})_p^{-1}$
proton-hole state is indicated as vertical thin line at $E_{\rm ex}-E_0=0$.
The threshold for the $(0s_{1/2})_p^{-1}$ hole state, which is the
excited state of the daughter nucleus, is at $E_{\rm ex}-E_0=18$ MeV
because of the different $S_p(j_p)$ in Eq.~(\ref{eq:Eex-E0}).

As we can see in the figure,
the
difference between the naive and mirror assignments is enhanced
in the ($\gamma$,p) reaction,
as
compared with the (d,$^3$He) case and, hence, the ($\gamma$,p) reaction is
more sensitive to the details of the $\eta$-nucleus interaction as we
expected. 
However, it might be difficult to
distinguish these two cases by experimental data, 
since the difference is seen only in the magnitude of the spectra above the
$\eta$ production threshold.
Hereafter, we
show only the results with the mirror assignment.

\begin{figure}[hbt]
\epsfxsize=12cm
\centerline{
\epsfbox{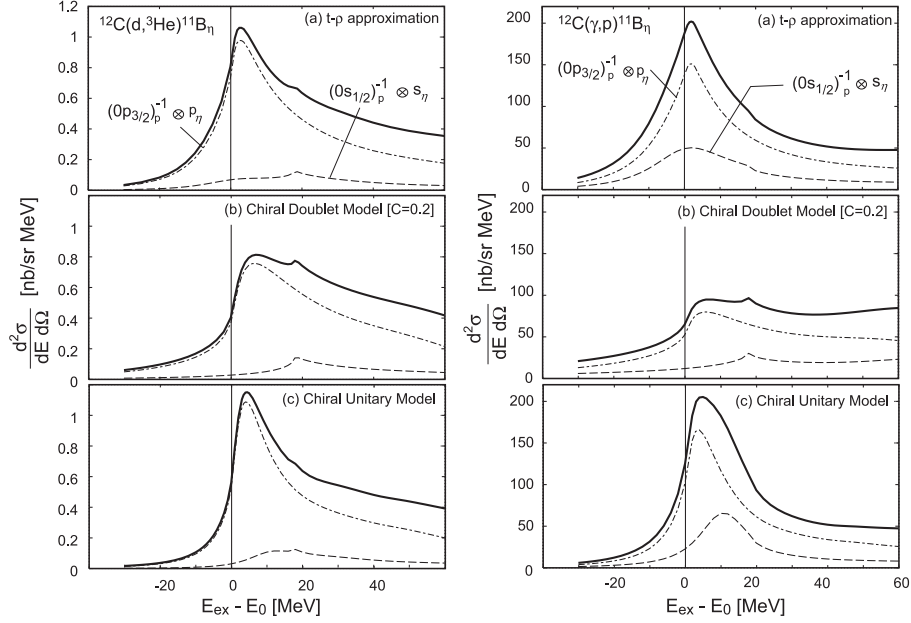}
}
\caption{Calculated spectra of $^{12}$C(d,$^3$He)$^{11}$B$\otimes \eta$
reactions at $T_{\rm d}=3.5$
 GeV in the left-side and
 $^{12}$C($\gamma$,p)$^{11}$B$\otimes \eta$ 
reactions at $E_\gamma=950$
 MeV in the right-side are shown as functions of the excitation
 energy $E_{\rm ex}$ defined in the text.
$E_0$ is the $\eta$ production threshold energy. The $\eta$-nucleus
interactions are evaluated
by (a) the t-$\rho$ 
 approximation, (b) the
 chiral doublet model with
 $C=0.2$
and (c) the chiral unitary model~\cite{PLB550}.
The total spectra are shown by the thick solid lines, and
the contributions from the
 dominant configurations
$(0s_{1/2})^{-1}_p \otimes s_\eta$ and
 $(0p_{3/2})^{-1}_p\otimes p_\eta$
are shown by dashed
 lines and dash-dotted lines, respectively, where the proton-hole states
 are indicated as $(n\ell_j)_p^{-1}$ and the $\eta$ states as $\ell_\eta$.
}
\label{fig:12C_eta}
\end{figure}

In Fig.~\ref{fig:12C_eta}, we show 
again the $^{12}$C target cases for the ($\gamma$,p) and (d,$^{3}$He) reactions
of the $\eta$ mesic nucleus, 
comparing the three different $\eta$-nucleus optical potentials. 
In Fig.~\ref{fig:12C_eta}(a), the spectra with the so-called
$t\rho$ optical potential are shown, which are calculated by putting $C=0.0$ in
the chiral doublet model. We show also the spectra obtained by the
chiral doublet model with $C=0.2$ in Fig.~\ref{fig:12C_eta}(b).
Comparing (a) and (b) in the left-side figure, we find that the repulsive nature
of the potential in the chiral doublet model with $C=0.2$ makes the bump
structure of the (d,$^3$He) spectrum broaden out to the higher energy
region, as discussed in Ref.~\cite{PRC66(02)045202}. 
On the right-hand side in Fig.~\ref{fig:12C_eta}, the same `broadening effect'
is seen also in the ($\gamma$,p) reaction.
We would like to stress here that 
this tendency to widen the spectra is seen more clearly in the
($\gamma$,p) spectra 
as a result of the transparency of the incident photon.
In Fig.~\ref{fig:12C_eta}(c), we show the spectra calculated
by the chiral unitary approach. We can see that the spectra of the
chiral unitary approach are shifted to the lower energy region like in
Figs.~\ref{fig:12C_eta}(a) as a result of the attractive
potential. It is very interesting to see which of spectra, (b) or (c), is
realized in experiment. 
It is also interesting to compare the spectra of the ($\gamma$,p) and
(d,$^{3}$He) reactions in experimental observation,
%
since we can expect to obtain the detail information of the
$\eta$-nucleus interaction by the comparison of the both data which have
different sensitivities to the nuclear surface and center.

As for the contributions from the bound $\eta$ states in the
($\gamma$,p) reaction, in Fig.~\ref{fig:12C_eta}(a) and (c), 
we find a certain enhancement in the $(0s_{1/2})_{p}^{-1}\otimes s_{\eta}$ configuration 
as a bump structure below the $0s$ state eta production threshold, which
is $E_{\rm ex}-E_{0}=18$ MeV. 
This is the indication of the $\eta$ meson bound in the $0s$ state, which is listed in
Table~\ref{tb:B.E.}.
Although the bound state is more clearly seen in the ($\gamma$,p) reaction, in
which we expect more reaction rates to pick up a proton in deeper states in the
target nucleus, as discussed before,
it is still hard to 
distinguish 
the bound state as a peak 
structure even
in 
the ($\gamma$,p) reaction due to the large width of the bound state and the larger contribution
of $(0p_{3/2})_p^{-1}$ state. 

\begin{figure}[hbt]
\epsfxsize=12cm
\centerline{
\epsfbox{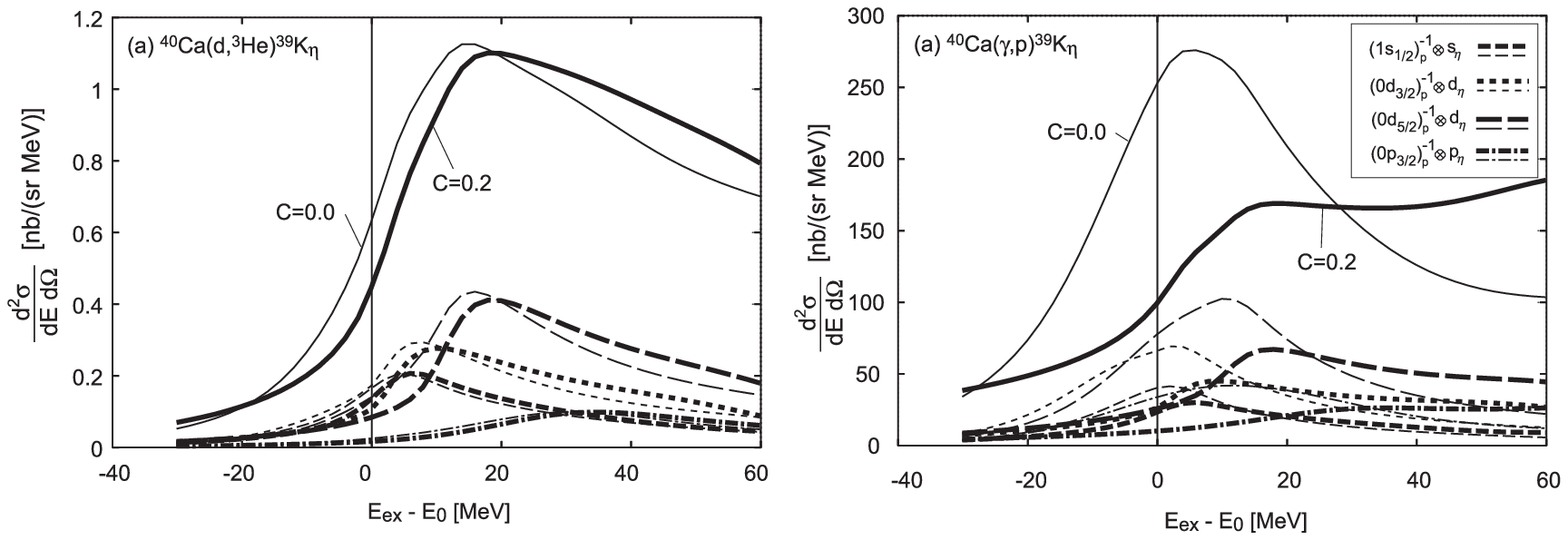}
}
\caption{
Calculated spectra for the formation of the $\eta$-mesic nuclei in
(a) the
 (d,$^3$He) reaction at $T_d=3.5$ GeV and (b) the ($\gamma$,p) reaction
 at $E_\gamma=950$ MeV are shown as functions of the excitation energy
 $E_{\rm ex}$ defined in the text.
$E_0$ is the $\eta$ production threshold energy. 
The $\eta$-nucleus interaction are evaluated by the chiral doublet model
 with $C=0.0$ and $C=0.2$, and the expected spectra are shown by thin
 lines and thick lines, respectively.
The total spectra are shown by the solid lines and dominant
 subcomponents 
 are shown
by dashed lines
as indicated in the figures.
Here, the proton-hole states
 are indicated as $(n\ell_j)_p^{-1}$ and the $\eta$ states as $\ell_\eta$.
}
\label{fig:40Ca_eta}
\end{figure}

Next we consider the heavier target $^{40}$Ca for the formation of the $\eta$
mesic nucleus. In this case, we have larger possibilities to have bound states
and to observe larger medium effects.
In Fig.~\ref{fig:40Ca_eta}, we show the calculated spectra of
$^{40}$Ca($\gamma$,p)$^{39}$K$\otimes\eta$ reaction using the chiral
doublet model with $C=0.0$ and $C=0.2$ together with those of the
(d,$^3$He) reaction for comparison.
%
%
As in the case of the $^{12}$C target,
we find larger difference between  the spectra with $C=0.0$ and
$C=0.2$ in the ($\gamma$,p) reaction than those in the (d,$^3$He)
reaction. 
In the (d,$^3$He) spectra, the total spectra are shifted around 5 MeV to
higher excitation energy region with almost same shape by changing the
$C$ parameter value from $C=0.0$ to $C=0.2$. On the other hand, the
($\gamma$,p) spectra are largely shifted to higher energy region with
significantly deformed shape.
This is the good advantage of the $(\gamma,p)$ reaction to deduce the information of the $\eta$ optical potential. 
%
We can also see the contribution from the bound
$\eta$ states for the $C=0.0$ case more clearly in the ($\gamma$,p) spectra.
The bound states were obtained in the $C=0.0$ case as energy 30.3 MeV
with width 42.5 in s-wave and energy 14.6 with width 50.7 in p-wave
\cite{PRC66(02)045202}. 
However,
as in the case of the (d,$^3$He) reaction,
it seems difficult to deduce
the contribution from a certain
subcomponent from the total
spectra even in the ($\gamma$,p) reaction
because of the large widths of the $\eta$ states.

In short summary for the $\eta$ mesic nucleus formation,
we find that
there exist certain discrepancies between the spectra obtained with the
different chiral models
and we also find that the ($\gamma$,p) spectra are more sensitive to
these discrepancies than the (d,$^3$He) spectra.

%

\subsection{$\omega$-nucleus system}
\label{sec:omega}
%
In the $\omega$ production in the ($\gamma$,p) reactions, the recoilless
condition is satisfied by the incident photon energy $E_\gamma=2.75$
GeV. The elementary cross section of the $\gamma+p\rightarrow p+\omega$
process is evaluated to be $0.3\mu$b/sr~\cite{PLB502}.  

\begin{figure}[hbt]
\epsfxsize=12cm
\centerline{
\epsfbox{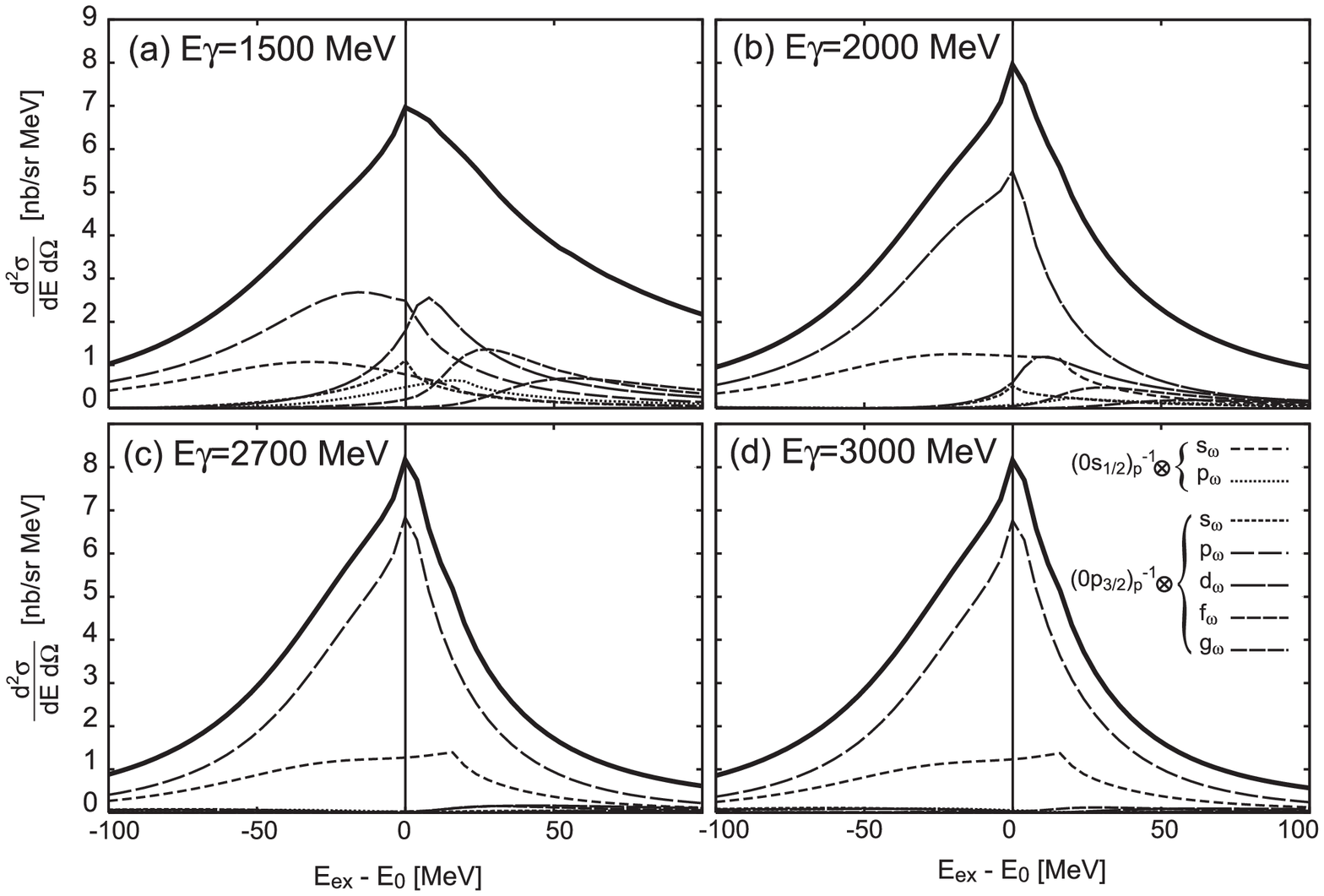}
}
\caption{
Calculated spectra of $^{12}$C($\gamma$,p) reaction for the formation
 of the $\omega$-$^{11}$B system 
with the attractive potential $V_\omega^{\rm (a)}$
at (a) $E_\gamma=1500$ MeV, (b)
 $E_\gamma=2000$ MeV, (c) $E_\gamma=2700$ MeV and (d) $E_\gamma=3000$
 MeV 
as functions of the excitation energy $E_{\rm ex}$ defined
 in the text. $E_0$ is the $\omega$ production threshold energy. The thick
 solid 
 lines represent the total spectra and the other lines
 represent
the contributions from the
dominant subcomponents as indicated in the figures.
}
\label{fig:12C_omega}
\end{figure}

First of all we show the result of the $\omega$ mesic nucleus formation spectra
with the optical potential $V_{\omega}^{\rm (a)}$ given in
Eq.~(\ref{eq:OPomega}).  
Shown in Fig.~\ref{fig:12C_omega} are the calculated spectra of the ($\gamma$,p)
reaction on the $^{12}$C target with various incident energies for $E_\gamma=1.5$, $2.0$, $2.7$ and $3.0$ GeV. So far the formation rate was calculated for the ideal 
kinematics to satisfy the recoilless condition. Here we study also the photon energy dependence of the expected spectra and its subcomponent contributions so as to see the experimental feasibility in the lower energy photon facilities. 
The energy dependence of the expected spectra is also useful for the
planned experiment at SPring-8 because it is required to use various
photon energy around the ideal kinematics to have the data with better
statistics~\cite{muramatsu}. 
%
%
As shown in Fig.~\ref{fig:12C_omega}, the spectra are dominated
by the contributions from the two configurations
$(0s_{1/2})_p^{-1}\otimes s_\omega$ and $(0p_{3/2})_p^{-1}\otimes p_\omega$
for the incident energies $E_\gamma=2.7$ GeV and $3.0$ GeV, which
roughly satisfy the recoilless condition. Even in this ideal case
for clear observations,
the bound state structure is hardly seen in the
spectra due to the large widths of the bound states.  
At $E_\gamma=2.0$ GeV, the total spectrum is still dominated by the two
configurations but other subcomponents have certain contributions above
the $\omega$ production threshold. For $E_\gamma=1.5$ GeV, we find that
many configurations contribute to the total spectrum and, 
%
hence, it is not easy to extract the contributions of specific configurations any more.
It is worth pointing out here that
the calculated results of the total spectra
in Fig.~\ref{fig:12C_omega} are similar for all cases shown here
insensitively to the photon energy. Thus, we need to keep in mind that the
contribution of each subcomponent has certain energy dependence even if
the total spectrum has almost no energy dependence. 

\begin{figure}[hbt]
\epsfxsize=12cm
\centerline{
\epsfbox{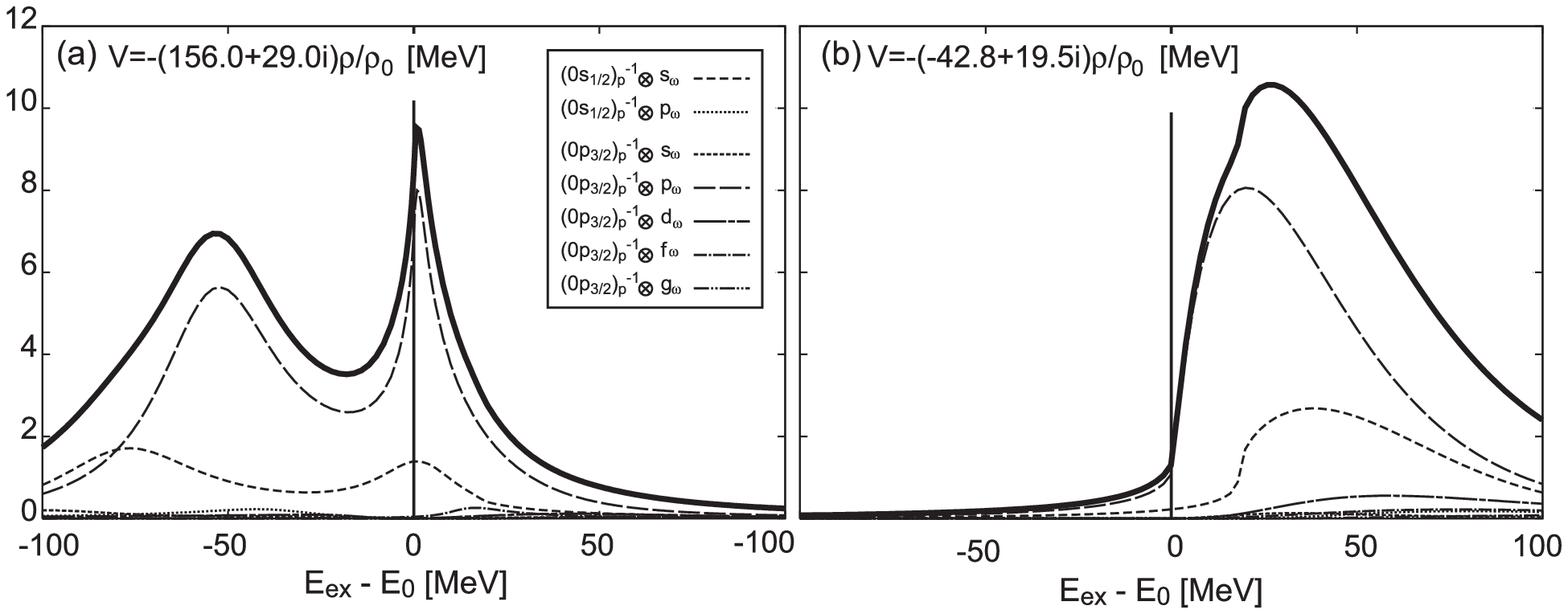}
}
\caption{
Formation spectra of the $\omega$ mesic nucleus in the ($\gamma$,p)
 reaction with the $^{12}$C target calculated 
 with the attractive potential $V_{\omega}^{\rm (b)}$ (a) and the
 repulsive potential $V_{\omega}^{\rm (c)}$ (b) at the photon energy
 $E_{\gamma}=2.7$ GeV.  $E_0$ is the $\omega$ 
 production threshold energy. The thick 
 solid 
 lines represent the total spectra and the other lines
 represent
the contributions from the
dominant subcomponents as indicated in the figures.}
\label{fig:12C_omega2}
\end{figure}

Next let us show the ($\gamma$,p) spectra calculated with the attractive
potential $V^{\rm (b)}_{\omega}$ and the repulsive potential $V^{\rm
(c)}_{\omega}$ defined in Eqs.~(\ref{eq:OPomega2})
(\ref{eq:OPomega3}). We evaluate the spectra with 
these potentials at the 
incident photon energy $E_{\gamma}=2.7$ GeV.  As seen in
Fig.~\ref{fig:12C_omega2}(a), in the case of the attractive potential
$V^{\rm (b)}_{\omega}$, the bound state structure is clearly seen thanks
to the narrow width of the bound state in $p$ state and dominant
contribution of the $(0p_{3/2})_p^{-1}\otimes p_\omega$
configuration. This is consistent with the ($\gamma$,p) spectrum shown
in Ref. \cite{PLB502}, in which the energy dependence on the $\omega$
optical potential is taken into account. Comparing these two spectra, we
see that the energy dependence of the potential
has relatively small contribution
to
fix
the global shape of the spectra and the energy dependence affects on the
reaction rate as a weak repulsion. In Fig.~\ref{fig:12C_omega2}(b), we
show the ($\gamma$,p) spectrum with the repulsive potential. Apparently
the shape of the spectrum is quite different from the previous two
attractive potentials. The repulsion of the omega meson inside the
nucleus makes the spectrum push to higher energies and almost nothing is
seen below the omega threshold.  
It would be interesting if one could measure the omega-nucleus system in
the ($\gamma$,p) reaction around the threshold of the omega production.

Here we have obtained three types of plots for the formation spectra of
the omega-nucleus system. In the first case, the optical potential
$V_{\omega}^{\rm (a)}$ has enough attraction to form bound states of the
$\omega$ meson in nucleus, but the widths of the bound states are
enlarged due to the strong absorption. Thus the whole spectrum of the
($\gamma$,p) reaction dose not show distinct peak structure caused by
the bound states even at the ideal recoilless kinematics, in which the
$(0p_{3/2})_p^{-1}\otimes p_\omega$ configuration dominants the
spectrum. In the second case, since the optical potential
$V_{\omega}^{\rm (b)}$ has more attraction and less absorption, a deeper
bound state of omega in $p$ wave is formed with a narrower width.  This
bound state is clearly seen in the spectrum at the recoilless
condition. In the third case, the optical potential $V_{\omega}^{\rm
(c)}$ is repulsive. The spectrum obtained with this potential has
definitely different from the above two cases. The missing mass
spectroscopy of the ($\gamma$,p) reaction in the omega production
energies is a good practical tool for the investigation of the
omega-nucleus interaction, at least it will be clearly seen if the
optical potential is attractive or repulsive.


\subsection{$\sigma$-nucleus system}
We propose the ($\gamma$,p) reaction as a new method to create the sigma
meson inside a nucleus, as discussed in Ref.~\cite{NPA710} with the
(d,$^{3}$He) reaction. In this subsection, we discuss the sigma meson
production in the $(\gamma, p)$ reaction on a heavy nucleus target
$^{208}$Pb. We show here the total spectral function formulated in
Sec.~\ref{sec:formulation} instead of the differential cross section
because of lack of our knowledge for the elementary cross section of the
sigma meson photoproduction. Since the total spectral function involves
the effective number which counts the reaction rate of the picked-up
proton induced by photon, we can investigate an impact of the narrow
bound states of the sigma meson around the two pion threshold, showing
the global shape of the spectral function without a definite number of
the cross section.  
The incident photon energy in the ($\gamma$,p) reaction is chosen to be
400 MeV so as to satisfy the recoilless condition for two-pion
production, since we are interested in the deeply bound sigma meson. 

%
%

\begin{figure}[hbt]
\epsfysize=4cm
\centerline{
\epsfbox{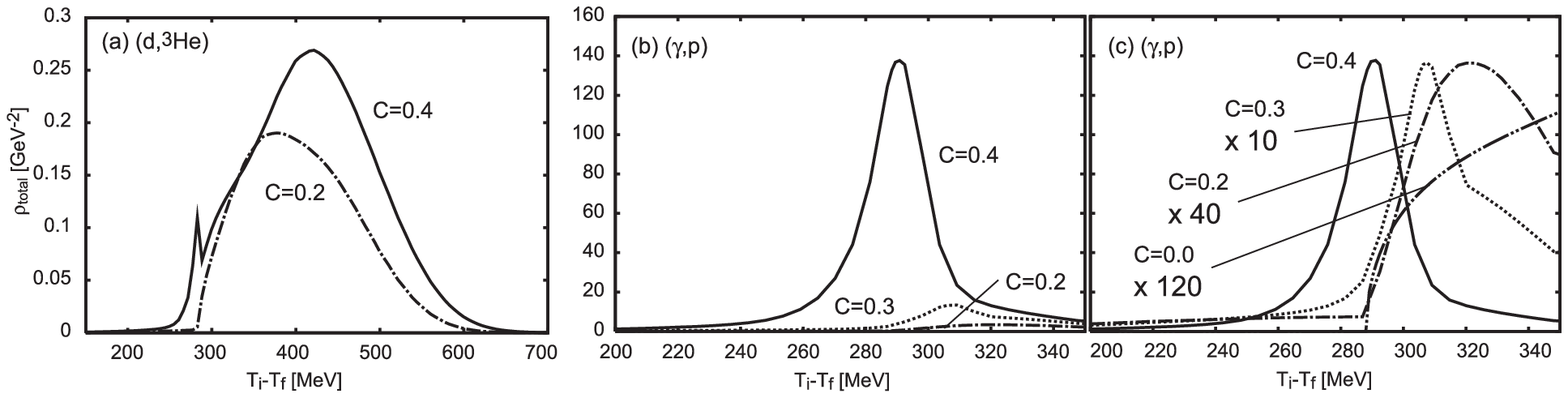}
}
\caption{
Total spectral functions $\rho_{\rm tot}$ defined in the text for 
(a) (d,$^3$He) reactions and (b) ($\gamma$,p) reactions on $^{208}$Pb target.
The incident particle energies $T_i$ are (a) 1.5 GeV and (b) 400
 MeV, respectively.
The solid lines indicate the spectral
 functions with the parameter $C=0.4$. The dashed and the
 dash-dotted lines represent the spectral functions with $C=0.3$ and
 $C=0.2$, respectively.
(c) Normalized spectral functions of the ($\gamma$,p) reactions.
The normalization factors are 10, 40, 120 for the
 $C=0.3$, $C=0.2$ and $C=0.0$ cases, respectively. The spectral function
 with $C=0.0$ reported Ref.~\cite{PRL82} is also shown 
 for comparison.}
\label{fig:tot_spe}
\end{figure}

Fist of all, we show the total spectral functions $\rho^{\rm
(d,^3He)}_{\rm tot}(\omega)$ for the (d,$^3$He) reactions with the
$^{208}$Pb target for comparison. The incident deuteron energy is chosen
to be 1.5 GeV in order to satisfy the recoilless condition  
around the emitted particle energy $T_f = T_i- 2m_\pi$.
Figure~\ref{fig:tot_spe}(a) shows the plots of $\rho^{\rm (d,^3He)}_{\rm
tot}(\omega)$ for the $C=0.2$ and $0.4$ cases. In the (d,$^3$He)
reaction, one finds a less prominent peak around $T_i-T_f= 2m_\pi$ for
$C=0.4$ as reported in Ref.~\cite{NPA710}.  
It would be hard, however, to distinguish such a small peak in
experiment. One sees also that the whole shape of the spectral function
for $C=0.4$ resembles that for 
$C=0.2$. This is because the large distortion effect of the hadronic
probe suppresses strongly production of the deeply bound states, which
have the significant enhancement around $\omega\sim 2m_\pi$, and, hence,
the total spectral function is dominated by the shallow bound states
with large widths like the $C=0.2$ case. 

%

Next, in  Fig.~\ref{fig:tot_spe}(b), we show the total spectral
functions for the ($\gamma$,p) reaction on the $^{208}$Pb target with
$C=0.2$, $0.3$ and $0.4$ as a function of the missing mass
$T_{i}-T_{f}$. 
As shown in the figure, the spectral function of the $C=0.4$ case is
significantly enhanced around $2m_\pi$ in a reflection of production of
the deeply bound states thanks to the transparency of the ($\gamma$,p)
reaction. The incident photon can go into the center of the nucleus,
knocking out a proton lying in a deep state, and, as a consequence, the
deeply bound sigma states with narrow widths are formed in the
nucleus. This is the good advantage of the photon induced reaction. 
The figure shows also a peak structure of the spectral function with
$C=0.3$ above the two-pion threshold. This is also the consequence of
production of the deeply bound sigma states. 

In order to compare the energy dependence of the total spectral
functions for the different strength of the chiral restoration, which is
parameterized as $C$, 
we show in Fig.~\ref{fig:tot_spe}(c) spectral functions normalized so as
to have the almost same peak height as the $C=0.4$ case. The normalization
factors are 10 for $C=0.3$, 40 for $C=0.2$ and
120 for $C=0.0$.  For comparison, we show also a spectral function with
$C=0.0$, in which the partial restoration of chiral symmetry does not
take place. The spectral function with $C=0.0$ is obtained in
Ref.~\cite{PRL82}, having the peak around $T_i-T_f\simeq 550$ MeV. Note
that the spectral function with $C=0.0$ is calculated without the
effective number and, therefore, the reaction mechanism is not involved
in the spectral function unlike the cases of $C\neq 0.0$. Nevertheless
the comparison gives us a schematic  view of the energy dependence. 

%
We can see the clear shift of the peak energy in this figure and can
expect that the shift could be a good signal of the chiral restoration
in the nuclear medium.


\subsection{Background Consideration}
So far we discuss the production rate of the mesons in nucleus in the
($\gamma$,p) reaction. In the missing mass spectroscopy of the emitting
proton, the production of the meson in the nucleus is expected to be
seen on top of a certain structureless background. In this section, we
roughly estimate the background of the ($\gamma$,p) reactions.  

First of all, in the $\eta$-nuclear system, the typical signal for the
creation of the eta meson in nucleus is estimated to be 100 nb/(sr MeV)
for the differential cross section of the $^{12}$C($\gamma$,p) reaction
at the recoilless energy $E_{\gamma}=950$ MeV as discussed in
Sec.~\ref{sec:eta}. 
On the other hand, experimental data~\cite{NPA415} showed that 
the differential cross section of the inclusive $^{12}$C($\gamma$,p)
reaction is $d^2\sigma/d\Omega dp {\aplt} 0.9 $ [$\mu$b/(sr MeV/c)] at
the photon energy $E_\gamma=580$ MeV and the emitted photon angle
$\theta_p=23^\circ$ in the laboratory frame. From the data, we estimate
the physical background due to the inclusive proton emission is
\begin{equation}
\frac{d^2\sigma}{d\Omega dE}=\frac{E}{p}\frac{d^2\sigma}{d\Omega dp}\sim 1.5 
\left[\frac{\mu{\rm b}}{\rm sr\ MeV}\right]  
\end{equation}
in the proton energy distribution.
%
%
%
Therefore, the signal over noise ratio
is expected to be around 1/15. However, the kinematical condition of
the ($\gamma$,p) reaction in Ref.~\cite{NPA415} is different from that
of the $\eta$-nucleus system formation reaction in Section
\ref{sec:eta}. Hence, we need to use theoretical models and/or more
appropriate experimental data to improve the background estimation.

As for the $\omega$ mesic nuclei formation by the ($\gamma$,p) reaction,
the signal is estimated to be 10 nb/(sr MeV) for the $^{12}$C target case in
Sec.\ref{sec:omega}, while the background is expected to be smaller than
100 nb/sr/MeV~\cite{muramatsu}. Thus, the signal over noise ratio is to
be around 1/10 from the calculated results.  

For the $\sigma$-nucleus systems, we do not have any background
estimations so far, since we cannot evaluate definite values for the
$\sigma$ meson production rates due to lack of the elementary cross
section. Conceivable physical backgrounds would be resonant enhancements
of the cross section due to nucleon excitations in nucleus. Especially
the kinematics considered here of the two pion production could conflict
with a $\Delta$ excitation in nucleus. A possible way to avoid such a
physical background is to observe more particles in the final state, for
instance, with taking coincidence of the emitted proton and two pion
with isospin 0.  Another possibility might be to observe the
($\gamma$,p) reactions with lighter nuclear targets and to see the mass
dependence of the peak around the two-pion threshold, since the deeply
bound sigma states will not be seen in a lighter nucleus. The detailed
discussions of the physical backgrounds are beyond the scope of the
present paper.

\section{Conclusion}
\label{sec5}
%
%
In this paper we have made a theoretical evaluation of the formation
rates of the $\eta$ and $\omega$ mesons in nuclei induced by the
($\gamma$,p) reactions in ideal recoilless kinematics. We have shown
the expected spectra in order to investigate the meson-nucleus
interactions. We have found that the ($\gamma$,p) reactions are good
practical tool to investigate the properties of the mesons created
deeply inside the nucleus due to the small distortion effects. This good
advantage provides the distinct difference in the formation spectra of
the $\eta$-nucleus system obtained by the two chiral models which are
based on the different physical pictures of the $N(1535)$ resonance. For
the $\omega$-nucleus system, we have compared three types of the
$\omega$ optical potentials in the ($\gamma$,p) spectra, showing the
definitely different shapes of the spectra.  

We have also investigate the ($\gamma$,p) spectra at the recoilless
condition for the two pion production in isoscalar channel
in order to study an impact of the
creation of the deeply bound states of the sigma meson associated with
the partial restoration of chiral symmetry in heavy nuclei. We have
found that a prominent enhancement around the two pion threshold in the
missing mass spectra in case of a sufficient strength of the partial
restoration in medium, owing to the transparency of the ($\gamma$,p)
reaction to create deeply bound states. Thus we expect that the
($\gamma$,p) reaction is a good tool to create the sigma meson in
nucleus. 

The study of the bound states is one of the
most promising method to investigate the meson properties at finite
density. 
Nevertheless the large natural widths of the meson bound states in
nucleus disable to distinguish contributions from each bound state. In
such a case, global conformation of the missing mass spectra is
necessary to extract valuable information of the meson nucleus
interaction. It is also beneficial to compare the spectra
of the ($\gamma$,p) and (d,$^{3}$He) reactions at corresponding
recoilless conditions, since each configuration differently contributes
to the total spectra due to the different distortion effects.   
We expect that the present results stimulate the experimental activities
and help the developments of this research field.

%
%

\section*{Acknowledgements}
We would like to thank T.~Hatsuda and T.~Kunihiro for fruitful
discussions on $\sigma$ meson properties in nuclear medium.
We acknowledge valuable discussions on photon induced reaction with
N. Muramatsu and H. Yamazaki.
We are also most grateful to M. Lutz for useful discussion on the
$\omega$-$N$ scattering amplitude. 
We also thank E.~Oset for his careful reading of our preprint and useful
comments. 
This work is partly supported by Grants-in-Aid for
Scientific Research of Monbukagakusho and Japan Society for 
the Promotion of Science (No. 16540254).
D.J.\ acknowledges support for his research work in Germany through a 
Research Fellowship of the Alexander von Humboldt Foundation.


\begin{thebibliography}{99}

\bibitem{Batty97}
For example; C.J. Batty, E. Friedman and A. Gal, Phys. Rep. {\bf 287}, 385
(1997).

\bibitem{NPA530etc}
H. Toki, S. Hirenzaki, T. Yamazaki, Nucl. Phys. {\bf A530}, 679 (1991),
S. Hirenzaki, H. Toki, T. Yamazaki, Phys. Rev. {\bf C44}, 2472 (1991).

\bibitem{PRC62}
H. Gilg {\it et al.}, Phys. Rev. C {\bf 62}, 025201 (2000);
K. Itahashi, {\it et al.}, {\it ibid.} {\bf 62}, 025202 (2000).

\bibitem{PRL88}
H. Geissel {\it et al.}, Phys. Rev. Lett. {\bf 88}, 122301 (2002).



\bibitem{PLB514}
P. Kienle, and T. Yamazaki, Phys. Lett. B {\bf 514}, 1 (2001);
H. Geissel, {\it et al.}, {\it ibid}. {\bf 549}, 64 (2002);
K. Suzuki, {\it et al.}, Phys. Rev. Lett. {\bf 92}, 072302 (2004).

\bibitem{PRC66(02)045202}
D. Jido, H. Nagahiro and S. Hirenzaki, Phys. Rev. C{\bf 66}, 045202
 (2002),
Nucl. Phys. {\bf A721}, 665c(2003).

\bibitem{PRC68(03)035205}
H. Nagahiro, D. Jido and S. Hirenzaki, Phys. Rev. C{\bf 68}, 035205 (2003).

\bibitem{PLB443etc}
K. Tsushima, D.H. Lu, A.W. Thomas, K. Saito, Phys. Lett. {\bf B443}, 26
 (1998),
K. Tsushima, D.H. Lu, A.W. Thomas, Phys. Rev. {\bf C59},1203 (1999).

\bibitem{EPJA6}
R.S. Hayano, S. Hirenzaki, A. Gillitzer,
Eur. Phys. J. A {\bf 6}, 99-105(1999).

\bibitem{PLB231}
M. Kohno and H. Tanabe, Phys. Lett. {\bf B} 231(1989) 219-223.

\bibitem{JPhysG}
A.I. Lebedev, V.A. Tryasuchev, J. Phys. G17, 1197 (1991).

\bibitem{PLB502}
E. Marco, W. Weise, Phys. Lett. {\bf B} 502(2001) 59-62.

\bibitem{PLB527}
S. Hirenzaki, E. Oset, Phys. Lett. B {\bf 527}, 69 (2002).

\bibitem{PR247etc}
See the reviews, T. Hatsuda and T. Kunihiro. Phys. Rep. {\bf 247}, 221
 (1994); G.E. Brown and M. Rho, {\it ibid.} {\bf 269}, 333 (1996).

\bibitem{PLB550}
C. Garcia-Recio, J. Nieves, T. Inoue and E. Oset, Phys. Lett. {\bf B550}, 47-54
(2002);
T. Inoue and E. Oset, Nucl. Phys. {\bf A710}, 354-370 (2002).

\bibitem{NPA624}
F. Klingl, N. Kaiser, W. Weise, Nucl. Phys. {\bf A624}, 527 (1997).

\bibitem{NPA650}
F. Klingl, T. Waas, W. Weise, Nucl. Phys. {\bf A650}, 299 (1999).

\bibitem{NPA710}
S. Hirenzaki, H. Nagahiro, T. Hatsuda, T. Kunihiro, Nucl. Phys. {\bf
 A710}, 131-144(2002).

\bibitem{APPB31etc}
E.E. Kolomeitsev, N. Kaiser, and W. Weise, Phys. Rev. Lett. {\bf 90},
 092501 (2003). 

\bibitem{PRC61}
S. Hirenzaki, Y. Okumura, H. Toki, E. Oset, and A. Ramos, Phys. Rev. C
 {\bf 61}, 055205 (2000).

\bibitem{NPA435}
O. Morimatsu, K. Yazaki, Nucl. Phys. {\bf A435}, 727(1985);
{\bf A483}, 493(1988).


\bibitem{ne0504010}
D. Trnka {\it et al.}, for the CBELSA/TAPS Collaboration, arXiv:nucl-ex/0504010.

\bibitem{ne0504016}
M. Naruki {\it et al.}, arXiv:nucl-ex/0504016.

\bibitem{muramatsu}
N. Muramatsu, private communication.

\bibitem{Lutz:2001mi}
M.~F.~M.~Lutz, G.~Wolf and B.~Friman,
Nucl.\ Phys.\ A {\bf 706} (2002) 431

\bibitem{PRD57etc}
S. Chiku, T. Hatsuda, Phys. Rev. {\bf D57}, R6 (1998),\\
S. Chiku, T. Hatsuda, Phys. Rev. {\bf D58}, 076001 (1998),\\
M.K. Volkov, E.A. Kuraev, D. Blaschke, G. Ropke, S.M. Schmidt,
 Phys. Lett. {\bf B424}, 235(1998).

\bibitem{PRL82}
T. Hatsuda, T. Kunihiro, H. Shimizu, Phys. Rev. Lett. {\bf
 82}, 2840 (1999).
\bibitem{PRD63}
D. Jido, T. Hatsuda, T. Kunihiro, Phys. Rev. {\bf D63}, 011901(R) (2000).

\bibitem{PRC18}
Y. K. Kwon and F. Tabakin, Phys. Rev. C {\bf 18}, 932 (1978).

\bibitem{PLB194}
S. Hirenzaki, T. Kajino, K.-I. Kubo, H. Toki, I. Tanihata,
 Phys. Lett. B {\bf 194}, 20 (1987).

\bibitem{HaiderLiu}
Q. Haider and L.C. Liu, Phys. Lett. B {\bf 172}, 257 (1986);
 Phys. Rev. C {\bf 34}, 1845(1986).

\bibitem{PRC44(91)738}
H. C. Chiang, E. Oset, and L. C. Liu, Phys. Rev. C{\bf 44}, 738 (1991).

\bibitem{PRL60(88)2595}
R.E. Chrien {\it et al.}, Phys. Rev. Lett. {\bf 60}, 2595 (1988).

\bibitem{Sokol}
G.A. Sokol {\it et al.}, Fizika B8, 85 (1999).

\bibitem{Pfeiffer}
M. Pfeiffer {\it et al.}, Phys. Rev. Lett. {\bf 92}, 252001 (2004).



\bibitem{PRD39(89)2805}
C. DeTar and T. Kunihiro. Phys. Rev. D{\bf 39}, 2805 (1989).

\bibitem{PTP106(01)873etc}
D. Jido, M. Oka, and A. Hosaka, Prog. Theor. Phys. {\bf 106}, 873
 (2001); D. Jido, Y. Nemoto, M. Oka, and A. Hosaka, Nucl. Phys. {\bf
 A671}, 471 (2000).

\bibitem{PRD57(98)4124}
Y. Nemoto, D. Jido, M. Oka, and A. Hosaka, Phys. Rev. {\bf D57}, 4124(1998).

\bibitem{NPA612}
N. Kaiser, T. Waas, W. Weise, Nucl. Phys. {\bf A612}, 297(1997).

\bibitem{Inoue}
T. Inoue, E. Oset, and M.J. Vicente Vacas, Phys. Rev. C{\bf 65}, 035204(2002).

\bibitem{NPA543}
M. Arima, K. Shimizu, and K. Yazaki, Nucl. Phys. {\bf A543}, 613 (1992).

\bibitem{PRC71}
A.M. Green and S. Wycech, Phys. Rev. C {\bf 71}, 014001 (2005).


\bibitem{PLB224(89)11}
T. Hatsuda and M. Prakash, Phys. Lett. B {\bf 224}, 11 (1989).

\bibitem{NPA640(98)77}
H. Kim, D. Jido, and M. Oka, Nucl. Phys. {\bf A640}, 77 (1998).

\bibitem{PLB362(95)23}
N. Kaiser, P.B. Siegel, W. Weise, Phys. Lett. {\bf B362}, 23 (1995).




\bibitem{PTP103}
Y. Umemoto, S. Hirenzaki, K. Kume, H. Toki, Prog. Theor. Phys. 103, 337
 (2000); Phys. Rev. C{\bf 62}, 024606 (2000).


\bibitem{ParticleData}
Review of Particle Physics, Particle Data Group; Phys. Lett. {\bf B}592,
 1, (2004).

\bibitem{JPSJ57}
S. Homma {\it et al.}, J. Phys. Soc. Jpn. 57, 828 (1988).

\bibitem{12C_ref}
S.L.Belostotskii et al., Sov.J.Nucl.Phys.41(6) 903 (1985).


\bibitem{40Ca_ref}
K. Nakamura, S. Hiramatsu, T. Kamae, and H. Muramatsu,
Phys. Rev. Lett. {\bf 33}, 853 (1974).


\bibitem{NPA415}
K. Baba {\it et al.}, Nucl. Phys. {\bf A415}, 462 (1984).









\end{thebibliography}
\end{document}